\documentclass[letterpaper,english,reprint,groupedaddress,amsmath,amssymb,aps]{revtex4}
\usepackage[T1]{fontenc}
\usepackage[latin9]{inputenc}
\usepackage{amsmath}
\usepackage{graphicx}
\usepackage{amssymb}

\makeatletter

\providecommand{\tabularnewline}{\\}

\@ifundefined{textcolor}{}
{%
 \definecolor{BLACK}{gray}{0}
 \definecolor{WHITE}{gray}{1}
 \definecolor{RED}{rgb}{1,0,0}
 \definecolor{GREEN}{rgb}{0,1,0}
 \definecolor{BLUE}{rgb}{0,0,1}
 \definecolor{CYAN}{cmyk}{1,0,0,0}
 \definecolor{MAGENTA}{cmyk}{0,1,0,0}
 \definecolor{YELLOW}{cmyk}{0,0,1,0}
 }

\usepackage{epsfig}\usepackage{amscd}\usepackage{caption}\usepackage{amstext}\usepackage{latexsym}

\usepackage{dcolumn}
\usepackage{bm}

\newcommand{\bra}[1]{{\langle{#1}}}
\newcommand{\ket}[1]{{\vert{#1}\rangle}}
\newcommand{\be}{\begin{equation}}
\newcommand{\ee}{\end{equation}}
\newcommand{\bqr}{\begin{eqnarray}}
\newcommand{\eqr}{\end{eqnarray}}

\makeatother

\usepackage{babel}

\makeatother

\usepackage{babel}

\makeatother

\usepackage{babel}

\begin{document}

\title{Microscopic and non-adiabatic Schrödinger equation derived from the
Generator Coordinate Method based on 0 and 2 quasiparticle HFB states}

\author{R.~Bernard}

\address{CEA, DAM, DIF, F-91297 Arpajon, France}

\author{H.~Goutte}

\address{Grand Accélérateur National d'Ions Lourds (GANIL), CEA/DSM-CNRS/IN2P3,
Bvd Henri Becquerel, 14076 Caen, France}

\author{D.~Gogny}

\address{Lawrence Livermore National Laboratory, Livermore, CA 94551, USA}

\author{W.~Younes}

\address{Lawrence Livermore National Laboratory, Livermore, CA 94551, USA}

\date{\today}
\begin{abstract}
A new approach called the Schrödinger Collective Intrinsic Model (SCIM)
has been developed to achieve a microscopic description of the coupling
between collective and intrinsic excitations. The derivation of the
SCIM proceeds in two steps. The first step is based on a generalization
of the symmetric moment expansion of the equations derived in the
framework of the Generator Coordinate Method (GCM), when both Hartree-Fock-Bogoliubov
(HFB) states and two-quasi-particle excitations are taken into account
as basis states. The second step consists in reducing the generalized
Hill and Wheeler equation to a simpler form to extract a Schrödinger-like
equation. The validity of the approach is discussed by means of results
obtained for the overlap kernel between HFB states and two-quasi-particle
excitations at different deformations. 
\end{abstract}


\maketitle

\section{Introduction}

The Generator Coordinate Method (GCM) is a very useful approach to
study large-amplitude collective modes. A review on the GCM and quantized
collective motion is given in reference \cite{Reinhard87}. It is
widely used for the study of ground state properties, and low-lying
collective states in even-even nuclei. For instance, in Ref. \cite{Sabbey07}
the Hill-Wheeler equation is solved for the axial quadrupole collective
coordinate, where the basis states are angular-momentum and particle-number
projected Hartree-Fock-Bogoliubov states using the Skyrme energy density
functional. Improvements have been made recently to generalize this
study by taking into account the triaxial degree of freedom, with
first results reported for $^{24}$Mg \cite{Bender08}. State of the
art configuration mixing calculations based on the GCM, including
variation after projection effects and projection of angular momentum
and particle number for the effective interaction D1S have also been
performed by the Madrid group \cite{Rodriguez07,Schwerdtfeger09}.
In Ref. \cite{Rodriguez07}, shell closures have been studied in the
$N=32,34$ mass regions and in \cite{Schwerdtfeger09} shape coexistence
near neutron number $N=20$ has been analyzed. When used with the
Gaussian Overlap Approximation (GOA) and employed with the full quadrupole
coordinate the GCM leads to a five dimensional collective Hamiltonian.
This model has been extensively put to the test recently in Ref. \cite{Bertsch07,Delaroche10},
where predictions for yrast states up to $6^{+}$, and non-yrast $0_{2}^{+}$,
$2_{2}^{+}$, $2_{3}^{+}$ states have been compared with experimental
data. A time-dependent version of the GCM has also been used to study
the fission process \cite{Berger84,Goutte05}. In these latter works,
the collective dynamics is derived from a time-dependent quantum-mechanical
formalism where the wave function of the system is of GCM form, and
for which a reduction of the GCM equation to a Schrödinger equation
has been made by means of the usual techniques based on the Gaussian
overlap approximation.

In all the aforementioned works the adiabatic assumption has been
made and no intrinsic excitations have been taken into account. Collective
vibrations have been considered to be completely decoupled from intrinsic
excitations. In the seventies, a few pioneering works were published,
in which the usual GCM was extended to include 2 quasi-particle (qp)
excitations (GCM+QP) \cite{Didong73,Muther77}. In Ref. \cite{Muther77},
the method was applied to $A\simeq50$ nuclei, for which excitation
spectra display both collective and non collective features. In fact,
calculations showed that vibrational degrees of freedom may be important
to describe the $2^{+}$ and $4^{+}$ states, while $6^{+}$ states
may be dominated by 2-qp components. In these calculations, the whole
pf shell was used as a basis assuming $^{40}\textrm{Ca}$ as an inert
core, and a modified version of the Kuo matrix elements was considered
for the residual interaction. A GCM+QP approach is well suited for
the study of many phenomena, for which the coupling between collective
and intrinsic states may play a role: i) shape coexistence where $K=0$
individual states could be coupled to low-lying vibrational collective
states, ii) backbending phenomena in rotational bands, iii) decay
of superdeformed states to normally deformed states and iv) non adiabatic
effects during the fission process.

Where the fully microscopic treatment of the whole fission process
from the initial stage of the fissioning system up to scission is
concerned, GCM based approaches are the best suited. They enable one
to treat the fission dynamics as a time-dependent evolution in a collective
space. Adiabatic calculations have shown that the dynamics is very
important for fission fragment distributions \cite{Goutte05}: it
is responsible for the large widths of the distributions. These results
have validated the adiabatic hypothesis made as a first approximation
for the description of low-energy fission. In fact, nuclear superfluidity
induced by pairing correlations, in addition to strongly influencing
the magnitude of the collective inertia, is responsible for a strong
rearrangement of the individual orbits with deformation. However,
there are some experimental indications that pairs are broken during
the fission process. For instance, manifestations of proton pair breaking
are observed in $^{238}\textrm{U}$ and $^{239}\textrm{U}$ nuclei
for an excitation energy of $2.3$ MeV above the barrier: first the
proton odd-even effect observed in the fragment mass distributions
decreases exponentially for an excitation energy slightly higher than
$2.3$ MeV \cite{Pomme93} and then the total kinetic energy drops
suddenly \cite{Pomme94,Vives00}. Some theoretical calculations have
studied the dissipation during the fission process, most of them are
based on a semi-classical formalism such as Focker-Plank equations
\cite{Nix84,Scheuter84,Kolomietz01}, or Hamiltonian equations with
one body dissipation and two-body viscosity \cite{Davies77,Carjan86},
or Langevin equations \cite{Nadtochy07,Borunov08}. Here we aim at
developing a microscopic non-adiabatic Schrödinger equation, in order
to obtain a microscopic description of the coupling between collective
and intrinsic excitations. We use in the following the abbreviation
SCIM which stands for Schrödinger Collective Intrinsic Model. The
newly developed SCIM formalism is presented in Section \ref{SCIM}.
In Section \ref{OVKER}, the calculation of the N-dimensional overlap
kernel between $0$ and $2$qp HFB states is shown, and the dependence
of its moments on the deformation is analyzed in details. Section
\ref{deterJ} is then devoted to the derivation of the inverse of
the overlap kernel. The Schrödinger equation is derived in Section
\ref{SE} and conclusions are drawn in Section \ref{concl}.

\section{Description of the formalism used to construct the SCIM \label{SCIM}}

Our derivation of the SCIM proceeds in two steps as described in the
two following paragraphs. The first step is based on a straightforward
generalization of the symmetric moment expansion of the equations
derived in the framework of the GCM \cite{Holzwarth72}. We refer
to the extensive review of the GCM given in \cite{RingSchuck80},
where most of the material and references related to this subject
are available. The second step consists in reducing the equation obtained
in section \ref{SME} to a simpler form which will be the starting
point to extract a Schrödinger-like equation.

\subsection{Generalization of the Symmetric Moment Expansion}

\label{SME}

In order to take into account the coupling between collective and
intrinsic degrees of freedom we use the ansatz

\begin{equation}
\ket{\Psi}=\int dqf_{0}(q)\ket{\Phi_{0}(q)}+\sum_{i\ne0}\int dqf_{i}(q)\ket{\Phi_{i}(q)}.\label{fog}\end{equation}
 In the present study, $\Phi_{0}(q)$ is the ground state at the $q$
value of the collective variable and $\Phi_{i}(q)$, with $i\ne0$,
are the associated 2-qp excitations whose precise definition and properties
are given in section \ref{sel2qp}. Following the usual procedure,
the weight functions $f_{i}(q)$ are found by requiring that the total
energy $\bra{\Psi}|\hat{H}\ket{\Psi}/\bra{\Psi}\ket{\Psi}$ calculated
with the function defined in Eq. (\ref{fog}) be stationary with respect
to arbitrary variation $\delta f_{i}^{*}$, which leads to the well
known Hill-Wheeler equation

\begin{equation}
\sum_{j}\int dq'\Bigr(H_{ij}(q,q')-EN_{ij}(q,q')\Bigl)f_{j}(q')=0\label{HW0}\end{equation}
 where the Hamiltonian and overlap kernels are defined in terms of
matrix elements such as \begin{eqnarray*}
H_{ij}(q,q') & = & \bra{\Phi_{i}(q')}|\hat{H}\ket{\Phi_{j}(q)},\\
N_{ij}(q,q') & = & \bra{\Phi_{i}(q')}\ket{\Phi_{j}(q)}.\end{eqnarray*}
 Eq. (\ref{HW0}) could be the starting point to derive a local expansion
by means of a Taylor expansion of the $f_{i}(q')$ around $q$, as
performed in reference \cite{Flocard75}. However we prefer here the
symmetric expansion \cite{Holzwarth72} which, in our point of view,
provides a natural way to expand on the non-locality. Its derivation
is very simple if we express the variational principle after performing
the change of the variables $\bar{q}=(q+q')/2$ and $s=q-q'$ in the
expression of the total energy. We only give here the resulting Hill-Wheeler
equation:

\begin{eqnarray}
\int dse^{isP/2}\Bigl[H(\bar{q}+\frac{s}{2},\bar{q}-\frac{s}{2})-EN(\bar{q}+\frac{s}{2},\bar{q}-\frac{s}{2})\Bigr]e^{isP/2}f(\bar{q})=0\label{HW1}\end{eqnarray}
 with \begin{equation}
P=i\frac{\partial}{\partial q}.\end{equation}
Let us mention that in all the equations of this formalism, we set
$\hbar=1$. At this stage it is important to mention that Eq. (\ref{HW1})
is obtained after performing successive integration by parts and consequently
the form given here supposes that the corresponding contributions
of the integrated terms (surface terms) vanish at the boundary of
the integration domain. Note that this is the case in all applications
of the GCM in spectroscopy studies \cite{Bertsch07,Delaroche10},
as well as in nuclear fission studies \cite{Berger84,Goutte05}. In
the following, we assume that we are in such a situation and pursue
our derivation with Eq. (\ref{HW1}) as given above. Now, the series
expansion of the exponentials is inserted in Eq. (\ref{HW1}) and
terms of the same order in $s$ are collected together. After integration
with respect to $s$ one is led to a symmetric moment expansion that
includes the coupling between collective and intrinsic variables.
Following the notations in reference \cite{RingSchuck80}, except
for the imaginary number \textquotedbl{}$i$\textquotedbl{}, the moments
of any operator $A$ is defined as:

\begin{equation}
A^{(n)}(\bar{q})=i^{n}\int_{-\infty}^{+\infty}dss^{n}A(\bar{q}+\frac{s}{2},\bar{q}-\frac{s}{2})\end{equation}
 and symmetric ordered products of operators $A^{(n)}(\bar{q})$ and
$P$ as \begin{equation}
\bigr[A^{(n)}(\bar{q})P\bigl]^{(n)}=\frac{1}{2^{n}}\sum_{k}C_{n}^{k}P^{n-k}A^{(n)}(\bar{q})P^{k}.\label{SOPO}\end{equation}
 With the procedure described above and the notations just given,
Eq. (\ref{HW1}) is transformed into \begin{equation}
\sum_{n}\frac{1}{n!}\Bigl(\bigr[H^{(n)}(\bar{q})P\bigl]^{(n)}-E\bigr[N^{(n)}(\bar{q})P\bigl]^{(n)}\Bigr)f(\bar{q})=0.\label{HW2}\end{equation}
Eq. (\ref{HW2}) is the compact form of a set of coupled equations
for the different components $f_{i}(q)$. Notice that the moments
occurring in its definition are square matrices whose dimension depends
on the number of excitations introduced in the description. It is
worth pointing out here that another difference with the usual approach
(no qp excitations) is that all moments, odd or even, must be included
in the summation. In order to give some more information about Eq.
(\ref{HW2}) we refer to Section \ref{OVKER} where we elaborate on
the properties of the selected GCM collective excitations in the present
work. There, it is shown that, with such a selection, the even moments
and odd moments of Hermitian operators are represented by real symmetric
matrices and imaginary anti-symmetric matrices respectively. As a
consequence the operators in Eq. (\ref{HW2}) are Hermitian. Finally,
since the Hamiltonian is time-reversal invariant, it is easy to check
that this operator is also invariant if collective and intrinsic coordinates
are time reversed as well. We will comment later on the effect this
result has on the Schrödinger equation we derive in Section \ref{SE}.

\subsection{Reduction of the Symmetric Moment Expansion to a Schrödinger like
equation}

Eq. (\ref{HW2}) is an infinite series which is an exact expansion
of the Hill-Wheeler equations in terms of local operators. It can
be transformed into a local Schrödinger equation by inverting the
expansion of the overlap kernel. The latter can be written formally
as \begin{equation}
\hat{N}(\bar{q})=\sum_{n}\frac{1}{n!}\bigl[N^{(n)}(\bar{q})P\bigl]^{(n)}=\left(\hat{N}^{1/2}(\bar{q})\right)^{+}\hat{N}^{1/2}(\bar{q}).\end{equation}
 In the following we use the definitions \begin{eqnarray}
\hat{N}^{1/2}(\bar{q}) & = & \hat{J}^{1/2}(\bar{q})\sqrt{N^{(0)}(\bar{q})}\nonumber \\
\left(\hat{N}^{1/2}(\bar{q})\right)^{+} & = & \sqrt{N^{(0)}(\bar{q})}\bigl(\hat{J}^{1/2}(\bar{q})\bigr)^{+},\label{kerN}\end{eqnarray}
 with \begin{eqnarray}
\hat{J}(\bar{q}) & = & I+\hat{u}(\bar{q})\nonumber \\
\hat{u}(\bar{q}) & = & \frac{1}{\sqrt{N^{(0)}(\bar{q})}}\sum_{n=1}\frac{1}{n!}\bigl[N^{(n)}(\bar{q})P\bigr]^{(n)}\frac{1}{\sqrt{N^{(0)}(\bar{q})}}.\label{defu}\end{eqnarray}
 These notations require further explanation. We use the property
that any Hermitian operator $A$ can be written in the form $A=S^{+}S$,
with an operator $S$ which is not necessarily Hermitian. Then, for
the sake of simplicity, the notation $S=A^{1/2}$ is used although
it is not strictly speaking the square root of $A$ unless $S^{+}=S$.\\
 We proceed further by defining a new set of collective wave functions,
$\{g_{i}(\bar{q})\}$, through the matrix relation, \begin{equation}
f(\bar{q})=\hat{N}^{-1/2}(\bar{q})g(\bar{q})=(N^{(0)}(\bar{q}))^{-1/2}\hat{J}^{-1/2}(\bar{q})g(\bar{q})\end{equation}
 to arrive finally from Eq. (\ref{HW2}) at the local Schrödinger
equation \begin{eqnarray}
\Biggl(\bigl(\hat{J}_{-1/2}(\bar{q})\bigr)^{+}\Bigl[\sum_{n}\frac{1}{n!}\frac{1}{\sqrt{N^{(0)}(\bar{q})}}\bigl[H^{(n)}(\bar{q})P\bigr]^{(n)}\frac{1}{\sqrt{N^{(0)}(\bar{q})}}\Bigr]\hat{J}_{-1/2}(\bar{q})-E\Biggr)g(\bar{q})=0,\label{eq:EqSch}\end{eqnarray}
 where we have defined the operator\begin{equation}
\hat{J}_{-1/2}(\bar{q})\equiv\hat{J}^{-1/2}(\bar{q}).\label{defJ-12}\end{equation}
 Eq. (\ref{eq:EqSch}) is formally the same as the one given in \cite{Flocard75}
but differs from it in many respects. The definitions of the moments
and the operator $\hat{u}(\bar{q})$ are not at all the same. Furthermore,
the operators occurring in its definition are represented by matrices
including, as already mentioned, those corresponding to odd moments.
Notice that the normalization \begin{equation}
\int d\bar{q}\bigl(g(\bar{q})\bigr)^{+}g(\bar{q})=1\end{equation}
 guarantees the normalization of the wave function defined in Eq.
(\ref{fog}). In this approach, the main difficulty is to determine
the operator $\hat{J}_{-1/2}(\bar{q})$, which, according to the definitions
given in Eq. (\ref{defu}), is solution of the equation \begin{equation}
\bigl(\hat{J}_{-1/2}(\bar{q})\bigr)^{+}(I+\hat{u}(\bar{q}))\hat{J}_{-1/2}(\bar{q})=I.\label{calcJ}\end{equation}
 In order to reduce its complexity, one is led to assume that the
series expansion representing the overlap kernel converges rapidly
and can be truncated after the second-order moment. That is to say,
we approach Eq. (\ref{calcJ}) by the simpler one \begin{eqnarray}
\bigl(\hat{J}_{-1/2}(\bar{q})\bigr)^{+}\Bigl[I+\frac{1}{\sqrt{N^{(0)}(\bar{q})}}(\bigl[N^{(1)}(\bar{q})P\bigr]^{(1)}+\frac{1}{2}\bigl[N^{(2)}(\bar{q})P\bigr]^{(2)})\frac{1}{\sqrt{N^{(0)}(\bar{q})}}\Bigr]\hat{J}_{-1/2}(\bar{q})=I.\label{calcJ2}\end{eqnarray}
 In section \ref{deterJ} we study Eq. (\ref{calcJ2}) in some detail
and explain how it can be solved with parametrizations of the form
\begin{equation}
\hat{J}_{-1/2}(\bar{q})=\sum_{n=0}^{4}\bigl[j_{(n)}(\bar{q})P\bigr]^{(n)}.\end{equation}
 The quantities $j_{(n)}(\bar{q})$ are unknown matrices which are
determined by solving Eq. (\ref{calcJ2}).

Once the inverse $\hat{J}_{-1/2}(\bar{q})$ is known it can be inserted
in Eq. (\ref{eq:EqSch}), and by means of the formulas given in the
Appendix \ref{A}, there is no difficulty in finding a general expansion
in terms of symmetric ordered products of operators. In other words
one can express Eq. (\ref{eq:EqSch}) in the form \begin{equation}
\sum_{n}(\bigl[S^{(n)}(\bar{q})P\bigr]^{(n)}-E)g(\bar{q})=0.\end{equation}
 Then a second order differential Schrödinger equation is derived
by limiting the summation to $n=2$. However, for obvious practical
reasons, one does not extract the $S^{(n)}(\bar{q})$ with the full
expansion of the Hamiltonian kernel but with a truncation similar
to the one introduced in the case of the overlap kernel. Thus, in
all that follows, our derivation of a Schrödinger equation relies
on the approximated expression

\begin{eqnarray}
\Biggl( & ( & \hat{J}_{-1/2}(\bar{q}))^{+}\frac{1}{\sqrt{N^{(0)}(\bar{q})}}(H^{(0)}(\bar{q})+\bigl[H^{(1)}(\bar{q})P\bigr]^{(1)}\nonumber \\
 & + & \frac{1}{2}\bigl[H^{(2)}(\bar{q})P\bigr]^{(2)})\frac{1}{\sqrt{N^{(0)}(\bar{q})}}\hat{J}_{-1/2}(\bar{q})-E\Biggr)g(\bar{q})=0,\label{eq:EqSch2-ord2}\end{eqnarray}
 with $\hat{J}_{-1/2}(\bar{q})$ satisfying Eq. (\ref{calcJ2}).

Such an approach assumes that the series expansions of the overlap
and Hamiltonian kernels converge rapidly. If the moments were independent
of $\bar{q}$ or if we could neglect their variations, Eq. (\ref{HW2})
would simply be a power expansion in the collective momentum $P$.
Then one could conclude naturally that this approach is limited to
the study of collective motions with momenta that are not too large,
or equivalently at moderate energies. Notice, that if we approximate
the kernels with a Gaussian \cite{RingSchuck80} of width $\sigma(\bar{q})$
the expansion (Eq. (\ref{HW2})) becomes a series with terms $[\sigma^{n}(\bar{q})P\bigr]^{(n)}$.
In that form it appears that a rapid convergence of Eq. (\ref{HW2})
requires that the collective wave functions, as well as the moments,
are slowly varying functions in a range given by the width of the
Gaussian. These arguments are only qualitative in the sense that they
do not give precise information regarding the energies up to which
one can expect the approximation to be reasonable. In relation to
this discussion we refer to the work by Bonche et al. \cite{Bonche90}
which contains a detailed comparison between two calculations of large
amplitude quadrupole dynamics in $^{194}$Hg. One is based on a direct
resolution of the Hill-Wheeler equation (HWE) as extracted with the
GCM, while the other uses a collective Schrödinger equation (CSE)
resulting from a local expansion of the HWE into a second order differential
equation. Although their approach to the CSE approximation is not
rigorously identical to the one described here, we think that one
can extract general deductions from their work. We will only quote
some of their conclusions here. The GCM and the CSE reproduced the
energies of most collective states fairly well. However one observes
significant differences in the collective wave functions at excitation
energies around $6$ MeV and above. More precisely the corresponding
HWE collective wave functions display more and more irregular structures
and rapid variations that a CSE approach is not able to reproduce.
We do not deny the fact that it is important to have a fair description
of the collective wave functions, but we do make the observation that
for excitations above $4$ or $5$ MeV one also has to worry about
their possible coupling with qp excitations, which is the main motivation
of this work.

At this stage one needs some more information on the kernel overlap
to proceed further in the reduction of the equations given above.
It is the purpose of the next section to present a numerical study
of this kernel in a wide range of deformations that would be encountered
in a microscopic description of the symmetric fission of $^{236}$U.

\section{The N-dimensional overlap kernel \label{OVKER}}

\subsection{Selection of the 2-qp generating wave function\label{sel2qp}}

In this paragraph, a set of generating wave functions is constructed
by means of the constrained Hartree-Fock-Bogoliubov method as described
in \cite{Berger84,Goutte05}. We limit this study to the case of one
generator coordinate $q$ associated to the total quadrupole deformation
of the even-even nucleus $^{236}$U. Furthermore, the calculations
are restricted to axial symmetry. As a consequence the qp are characterized
by their projection $K$ of the total angular momentum on the symmetry
axis. We denote the ground state (or Bogoliubov vacuum) at deformation
$q$ by $\ket{\Phi_{0}(q)}$ and recall here its definition in terms
of the qp destruction operators $\eta_{i}(q)$ and their time reversal
conjugate $\eta_{\bar{i}}(q)=\hat{T}^{+}\eta_{i}(q)\hat{T}$, where
$\hat{T}$ is the time reversal operator: \begin{equation}
\ket{\Phi_{0}(q)}=\prod_{i>0}\eta_{i}(q)\eta_{\bar{i}}(q)\ket{0}.\end{equation}
 Furthermore we restrict our calculations by introducing another symmetry
which is associated with the invariance of the Bogoliubov fields when
performing a reflexion with respect to the $x0z$ plane. In that case
the parity $\pi$ is an additional quantum number and the qp operators
are then characterized by $k\equiv\{n_{k},\pi_{k},K_{k}\}$ where
$n_{k}$ is an index which distinguishes single qp states inside the
same block $\{\pi_{k},K_{k}\}$. At each deformation we can then define
a set of 2-qp excitations \begin{equation}
\eta_{i_{1}}^{+}(q)\eta_{i_{2}}^{+}(q)\ket{\Phi_{0}(q)}.\end{equation}
 This study is limited to the description of the coupling of the ground
state with 2-qp excitations. Consequently, since the Hill-Wheeler
equation conserves the total projection of the angular momentum on
the symmetry axis $(z)$ and since the ground state is a state $K=0$,
it is clear that the quantum number of 2-qp excited states must satisfy
$K_{i_{1}}+K_{i_{2}}=0$. In addition, by using similar arguments
with the time reversal transformation, we arrive at the conclusion
that the qp excitations must be symmetric under time reversal. Thus
the $N$ selected configurations at deformation $q$ are written in
the form: \begin{equation}
\begin{cases}
\ket{\Phi_{0}(q)}\\
\ket{\Phi_{i}(q)}=\frac{1}{\sqrt{2}}[1+\delta_{i_{1}i_{2}}(\frac{1}{\sqrt{2}}-1)]\bigl(\eta_{i_{1}}^{+}(q)\eta_{\bar{i}_{2}}^{+}(q)\\
\hspace{2cm}+\eta_{i_{2}}^{+}(q)\eta_{\bar{i}_{1}}^{+}(q)\bigr)\ket{\Phi_{0}(q)},\,\,\, i=1,N-1.\label{defHFBi}\end{cases}\end{equation}
 Notice that for the 2-qp states, the index $i$ denotes the couple
$(i_{1},i_{2})$. For simplicity of notation we have not added another
index to indicate whether the excited configurations are built with
two neutrons or two protons qp. Finally, for sake of convenience and
since the 2-qps $i_{1},i_{2}$ are in the same block $\{\pi_{i},K_{i}\}\equiv\{\pi_{i_{1(2)}},K_{i_{1(2)}}\}$,
we also use in the following the notation $i=\{K_{i}^{\pi_{i}},(a_{i})\}$
where the letter $a_{i}$ which distinguishes the couple $(n_{i_{1}},n_{i_{2}})$
in the same block $\{\pi_{i},K_{i}\}$ is used if necessary. These
wave functions which are not eigenfunctions of the proton and neutron
number operators $\hat{N}_{p},\hat{N}_{n}$ should in principle be
projected on the subspace of states with good particle numbers. The
expression of the overlap kernel is readily expressed with the usual
techniques to calculate the overlap between two Bogoliubov vacuum
\cite{Haider92}. The difficulty in projecting is essentially numerical
since the projection method requires repeating the calculation of
the overlap at different angles in order to calculate an integral
with a discretization method \cite{Anguiano01}. Such calculation,
if feasible in the study proposed here, would represent a considerable
amount of computational work and numerical studies. At this stage
it is worth pointing out that a description of the fission process
from the first barrier to scission involves a wide range of deformations
typically in the interval $[50\,\textrm{b},550\,\textrm{b}]$. Furthermore,
an accurate calculation of the potential energy surface, in particular
close to the scission point, necessitates the use of a very large
dimensional space of harmonic oscillator basis states. As an example,
we indicate that in a one-center basis the size of the $K=1/2$ block
is $[150\times150]$. In view of these remarks, we find some justification
in presenting this formalism with unprojected wave functions. We stress
that until now our formalism does not depend on the particular model
used to calculate the kernels. Finally, let us conclude this section
with a different issue which is still related to this question of
particle number. It concerns the restoration of the correct average
values of the number of protons and neutrons of the Hill-Wheeler solutions.
This problem is studied in details in reference \cite{Bonche90} where
the authors introduce in the Hill-Wheeler equation the kernels calculated
with the two operators $\hat{N}_{p}$, $\hat{N}_{n}$ and their associated
Lagrange parameters. The implementation of this method in our formalism
is straightforward.

\subsection{Calculation of the overlap kernel\label{calcovekr}}

\subsubsection{Generality \label{gene}}

Calculations presented in this paper correspond to different excitations
along the symmetric fission barrier in $^{236}$U. The HFB equations
are solved by expanding the single-particle states onto an axial harmonic
oscillator (HO) basis. A one-center basis with $N=14$ major shells
has been used. The calculations are performed with the Hartree-Fock
+ BCS method and a constraint on the quadrupole operator. They have
been restricted to axial and left-right symmetries. The full Bogoliubov
approach has not been considered here for reasons of computational
time only. Because of the symmetries imposed in this calculation,
the Bogoliubov or (HF+BCS) transformation can be chosen to be real.
As a result all quantities occurring in the formalism are real. This
is the case in particular of the overlap matrix which is denoted by
\begin{equation}
N_{ij}(q,q')=\bra{\Phi_{i}(q)}\ket{\Phi_{j}(q')}.\end{equation}
 Calculations of this matrix have been performed at different elongations
along the symmetric barrier. The formalism used to extract these matrix
elements is given in detail in Appendix \ref{B}. In all what follows
the results are presented in the variables $\bar{q}$ and $s$ defined
in the first paragraph of Section \ref{SCIM}. More precisely at each
$\bar{q}$ given in barns, we plot the matrix elements $N_{ij}(\bar{q}+s/2,q-s/2)$
as function of $s$ running in intervals wide enough that the overlaps
$N_{ij}(\bar{q}+s/2,q-s/2)$ drop to zero. The $\{\bar{q}\}$ have
been chosen in the range $[20\,\textrm{b},350\,\textrm{b}]$ with
a step of $10\,\textrm{b}$ up to $\bar{q}=100\,\textrm{b}$ and a
step of $25\,\textrm{b}$ beyond. Let us mention that deformations
close to $160\,\textrm{b}$, namely $\bar{q}=150\,\textrm{b}$ and
$\bar{q}=175\,\textrm{b}$, are not discussed here, since they correspond
to a junction between two valleys having different hexadecapole deformations.
Such a study requires to treat explicitly the hexadecapole degree
of freedom \cite{Dubray08,Younes09}. 

\begin{figure*}[t]

\begin{centering}
\includegraphics[width=8cm]{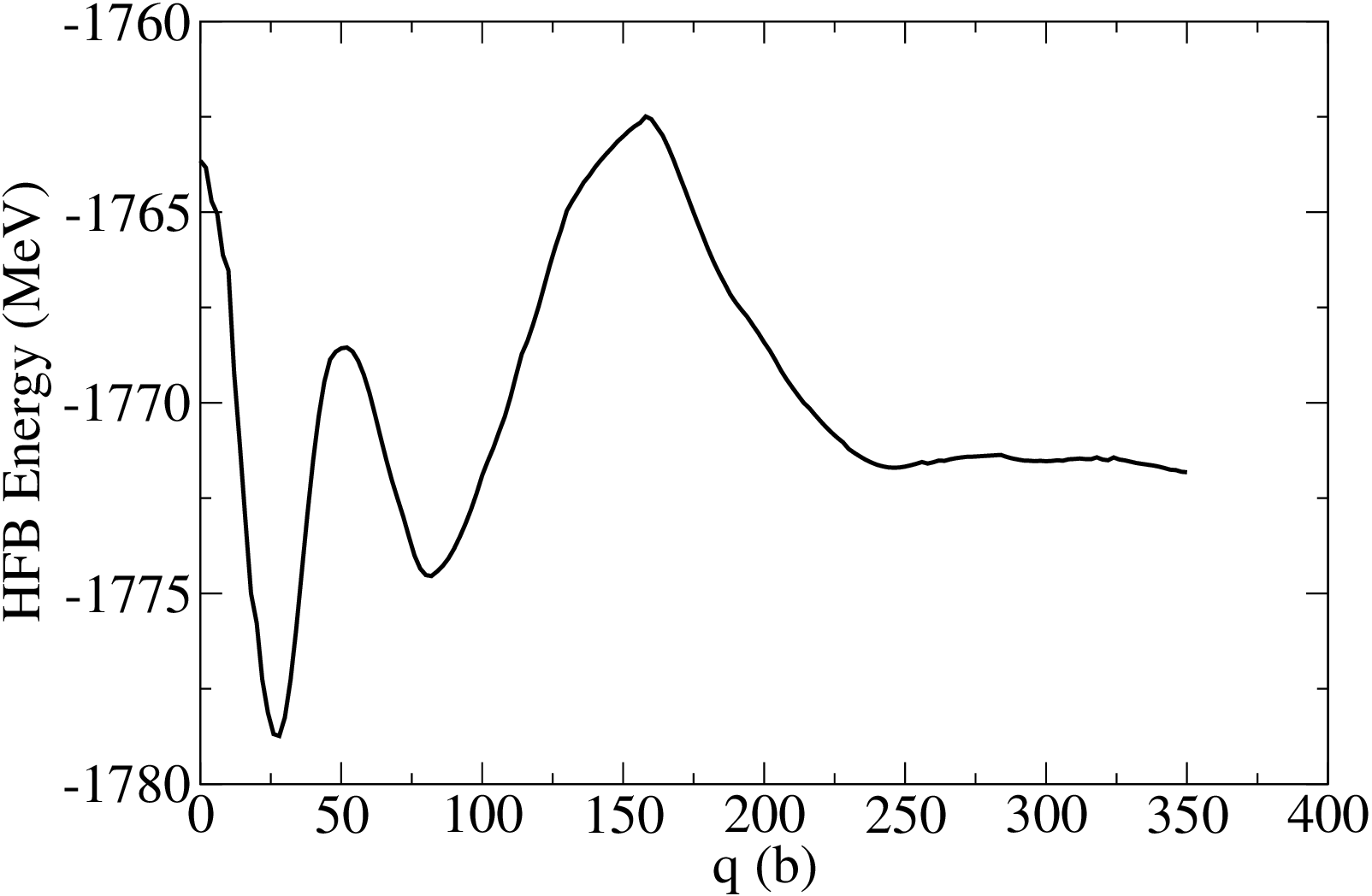} \caption{\label{bar_sym} HFB potential energy along the symmetric fission
barrier in $^{236}$U.}

\par\end{centering}

\centering{}%
\end{figure*}

The symmetric barrier is plotted in Fig. \ref{bar_sym}. The ground
state is located near $30\,\textrm{b}$, the first barrier at $50\,\textrm{b}$,
the second well at $80\,\textrm{b}$ and the second saddle point at
$160\,\textrm{b}$.

The intrinsic levels chosen to construct the 2-qp excitations are
the lowest-energy proton $K^{\pi}=1/2^{-},1/2^{+},3/2^{-},3/2^{+},...$
qp levels. Among them, the selected intrinsic excitations are those
which minimize the deviation from the average proton and neutron numbers.
Results are presented in the present work only for proton excitations
and similar results and conclusions are expected for neutron excitations.

Some comments are in order concerning phase problems that we encountered
in the construction of these 2-qp excitations. In fact the qp states
or operators $\eta_{i}$ which result from the diagonalization of
the Bogoliubov Hamiltonian are defined up to an arbitrary phase which
in the present case is real, i.e. $\pm1$. The phases of two conjugate
qps $\eta_{\bar{i}},\eta_{i}$ are then the same by construction and
consequently the Bogoliubov vacuum defined above does not depend on
these phases. The same conclusion applies to the case of two conjugate
qp excitations $\ket{\Phi_{i}(q)}=\eta_{i_{1}}^{+}\eta_{\bar{i}_{1}}^{+}\ket{\Phi(q)}$
but it is not the case for all other $2$qp excitations. It is worth
stressing that if we were using the qp as they come out of the diagonalization
of the Bogoliubov Hamiltonian, the corresponding non-diagonal matrix
elements of the kernels can become discontinuous at random as function
of the deformation, which would prevent the extension of the GCM with
those excitations. In order to solve this problem, the single-qp excitations
used in the construction of the 2-qp excitations have been followed
continuously as a function of deformation. This is achieved with the
following simple procedure. Any single-qp state $\ket{i,q+\delta q}$
at deformation $q+\delta q$ is associated with $\ket{i,q}$ by maximizing
the single-particle overlap $|\bra{i,q}\ket{k,q+\delta q}|$ among
all $k$ in the $1$-qp spectrum at deformation $q+\delta q$. This
overlap is nothing but the anti-commutator $\{\eta_{i}^{+}(q),\eta_{k}(q+\delta q)\}$.
Denoting by $k$ the qp which achieves this maximization, we set $\ket{i,q+\delta q}\equiv\varphi_{i}(q)\ket{k,q+\delta q}$
where a phase given by \begin{equation}
\varphi_{i}(q)=\frac{\bra{i,q}\ket{k,q+\delta q}}{|\bra{i,q}\ket{k,q+\delta q}|}\end{equation}
 has been introduced. It is clear that with such procedure the matrix
elements of the overlap kernel become continuous functions of the
deformation. Let us mention that the problem of the sign of the overlap
has been recently addressed by L. Robledo in \cite{Robledo08} for
general HFB wave functions that do not have any spatial symmetry (triaxial)
and also breaking of the time reversal symmetry.

\begin{figure}[!h]

\begin{centering}
\includegraphics[width=8cm]{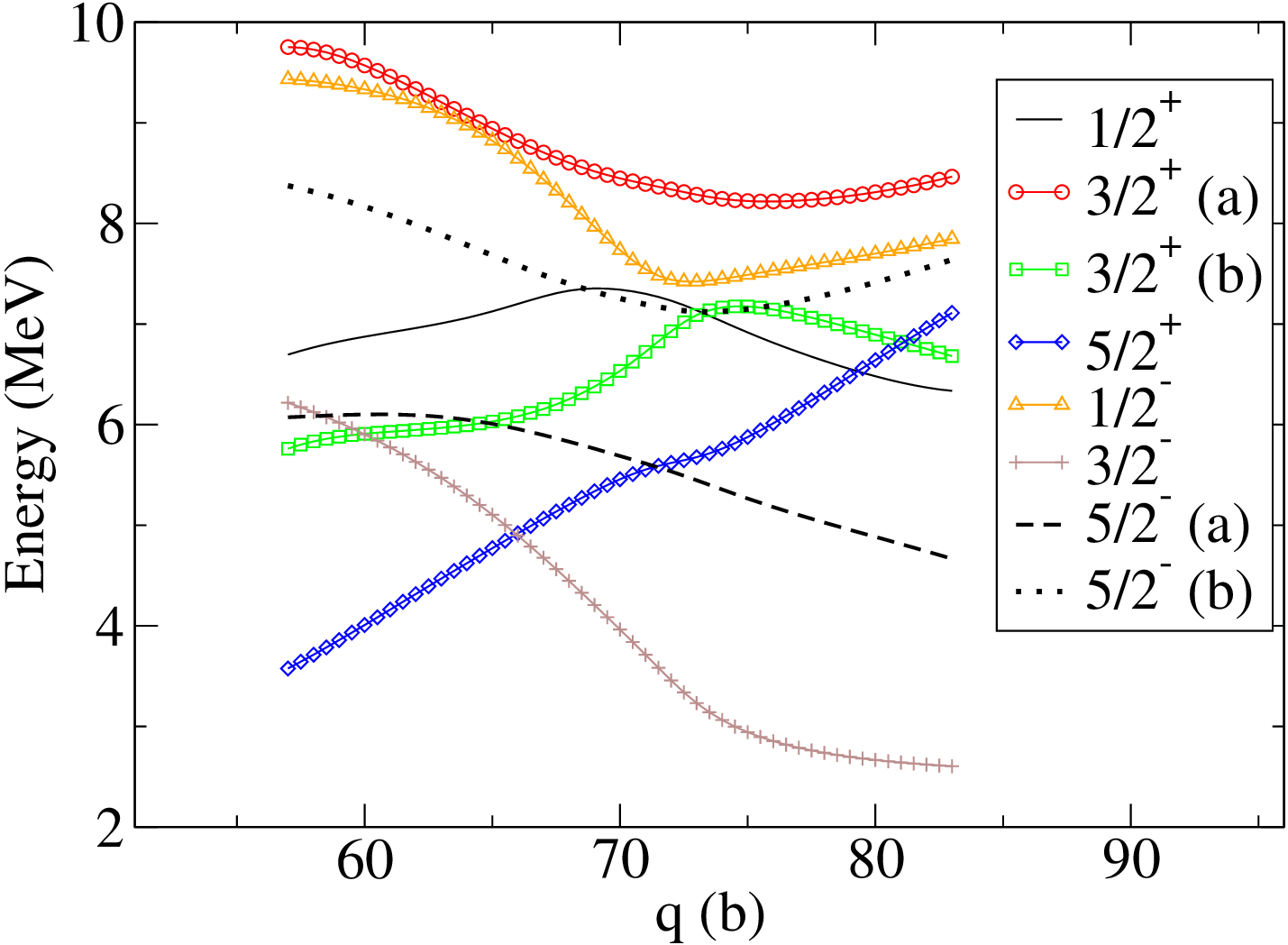}
\caption{\label{2qpcont} Energy of the selected 2-qp excitations around $q=70\,\textrm{b}$,
as a function of deformation. Letters are used to differentiate excitations
with same $K^{\pi}$.}

\par\end{centering}

\centering{}%
\end{figure}

The evolution of the energy of 2-qp excitations is plotted in Fig.
\ref{2qpcont} as a function of the deformation around $\bar{q}=70\,\textrm{b}$.
It shows that the lowest excitation is located around $3$ MeV. 
 No significant difference is observed between proton and neutron
excitations. 

Let us mention that we need not limit ourselves to excitations that
conserve the average number of particles, but as a result the number
of excitations increases rapidly when we allow the mean particle numbers
$\bar{N}_{i}$ and $\bar{Z}_{i}$ to vary by 1 particle around their
average value. For instance for an energy below only $5$ MeV and
with a mean particle number in the range $92-1\le\bar{Z}_{i}\le92+1$,
$144-1\le\bar{N}_{i}\le144+1$, we already find $6.5$ proton 2-qp
excitations and $11.5$ neutrons per deformation.


%

\begin{figure}[!h]

\begin{centering}
\includegraphics[width=8cm]{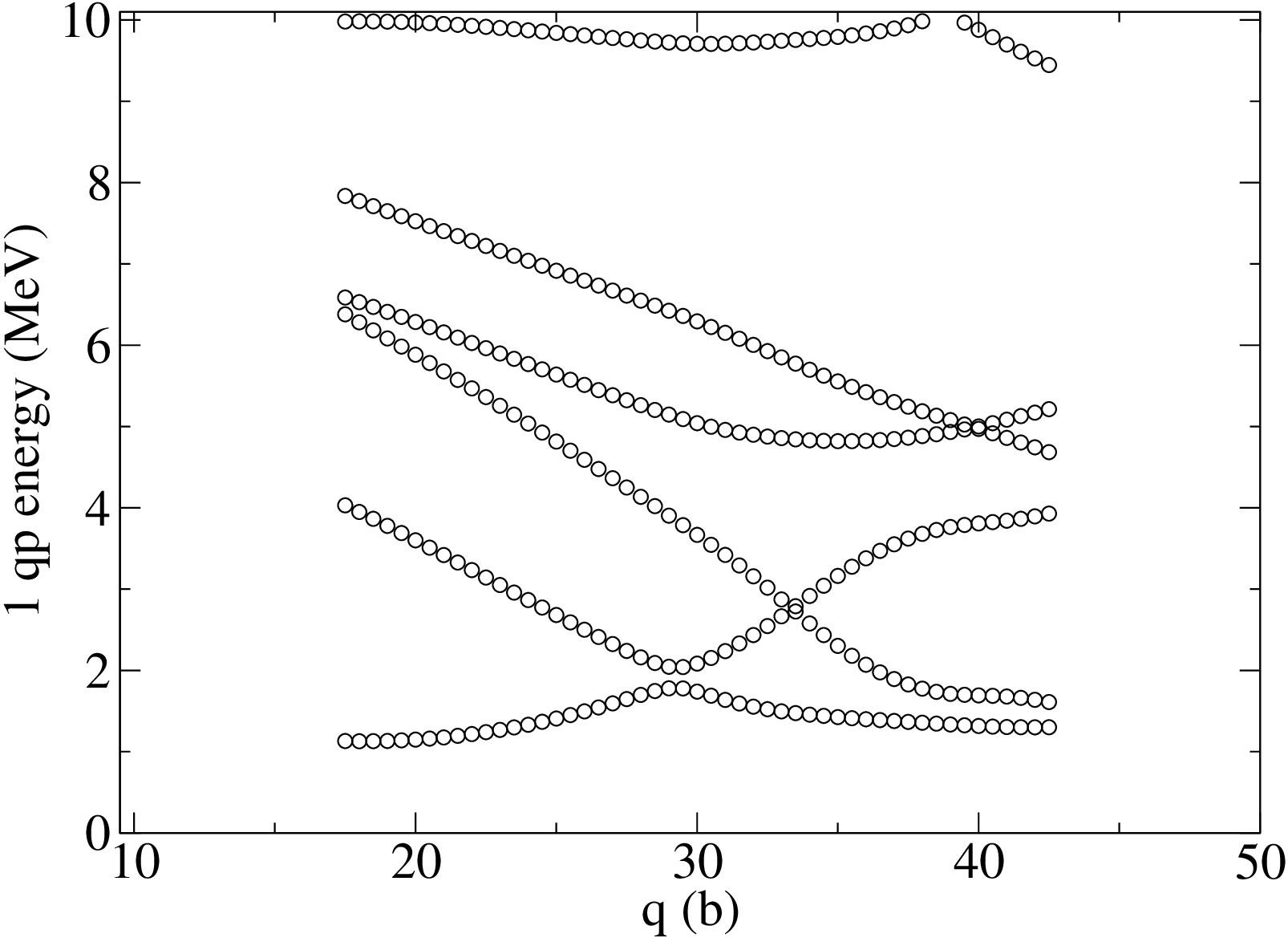} 
\caption{\label{schemaqp} Energy of the single-qp excitations around $q=30\,\textrm{b}$
for $K^{\pi}=1/2^{+}$.}

\par\end{centering}

\centering{}%
\end{figure}

As we follow the single-qp states by continuity as a function of deformation,
levels with the same $K$ and $\pi$ quantum numbers can occasionally
cross or be pushed back from each other. Such features are common
and can occur anywhere along the deformation. Such cases are illustrated
in Fig. \ref{schemaqp} around $q=30\,\textrm{b}$ for $K^{\pi}=1/2^{+}$.
The figure shows that a repulsion occurs between the two first levels
at a deformation $q_{r}=29.5\,\textrm{b}$, defined as the deformation
where the difference in energy between the 2 levels is minimal, whereas
crossings are observed at $q=34\,\textrm{b}$ and $q=40\,\textrm{b}$.
Note that, unlike crossings, repulsions correspond to a mixing of
the two incoming levels.

\subsection{Numerical results for the matrices $N_{ij}(\bar{q}+s/2,\bar{q}-s/2)$
of the overlap kernel \label{num_ove}}

As already mentioned in \cite{Bonche90}, an accurate calculation
of the complex behavior of the overlap kernel requires a careful determination
of the HFB states. The overlap kernels are the determinants of the
matrices built from the overlaps of all pairs of individual orbits:
they are very sensitive to the details of those orbits. For this reason
one must achieve a very accurate convergence of the HFB wave functions
if one wants to get reliable values of the overlap kernels. As a consequence,
all the HFB states here have been obtained with a very high degree
of accuracy. More precisely, a convergence of $10^{-6}\,\textrm{f}\textrm{m}^{-3}$
has been achieved on the generalized density matrix. Let us note that
the parameters of the Harmonic Oscillators used to develop the HFB
states have been optimized in each region at the deformation $\bar{q}$
(i.e. $q=q'$), and kept constant for the neighboring deformations.
For reference we have thought it interesting to show the plot of these
matrices as function of $s$ for different values of $\bar{q}$ even
though the quantities which count in this formalism are their moments.
The latter are presented in section \ref{anamom}.

\subsubsection{Calculation of the diagonal matrices $N_{ii}(\bar{q}+s/2,\bar{q}-s/2)$}

\begin{figure}[!h]

\begin{centering}
\includegraphics[width=7cm]{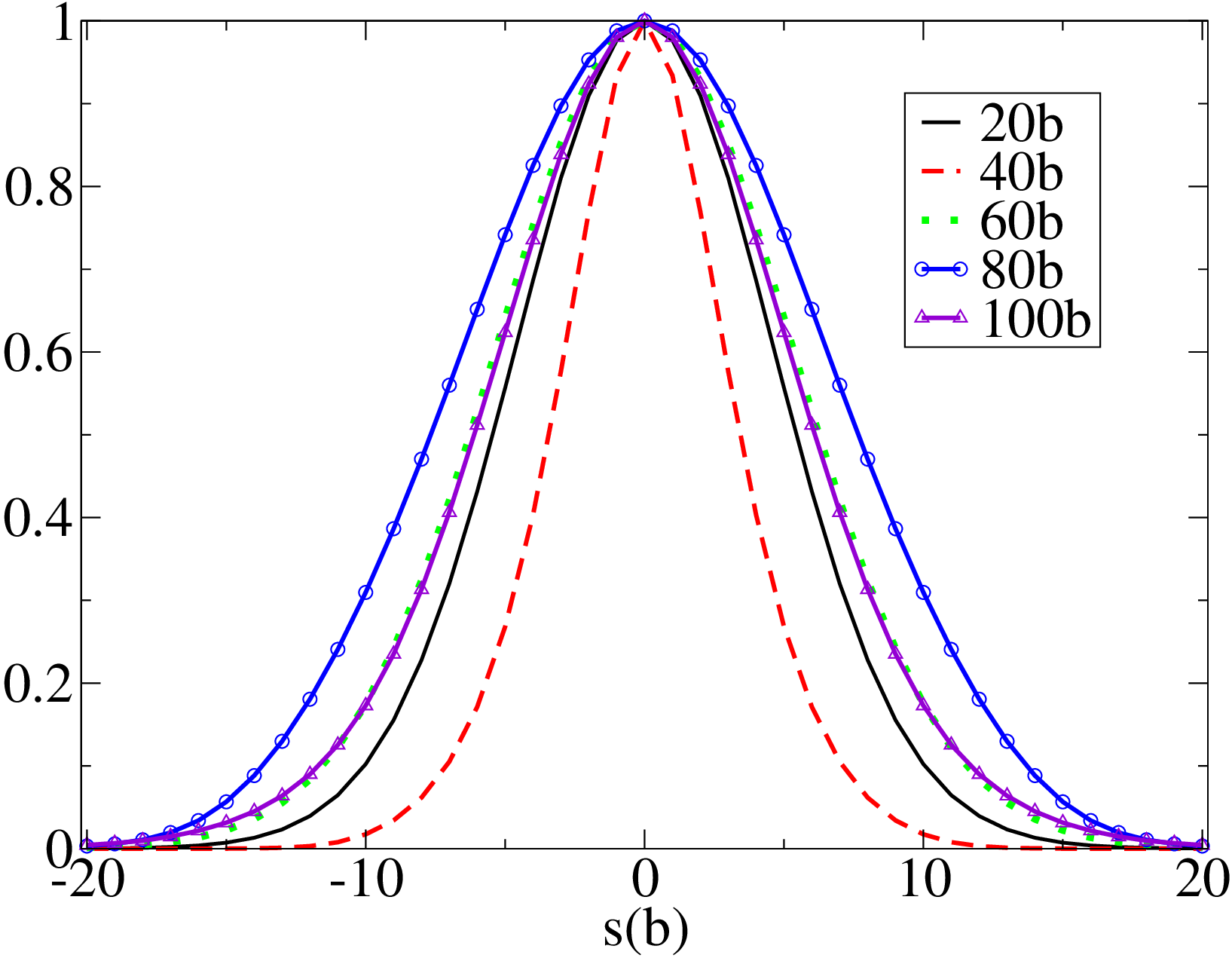} 
\includegraphics[width=7cm]{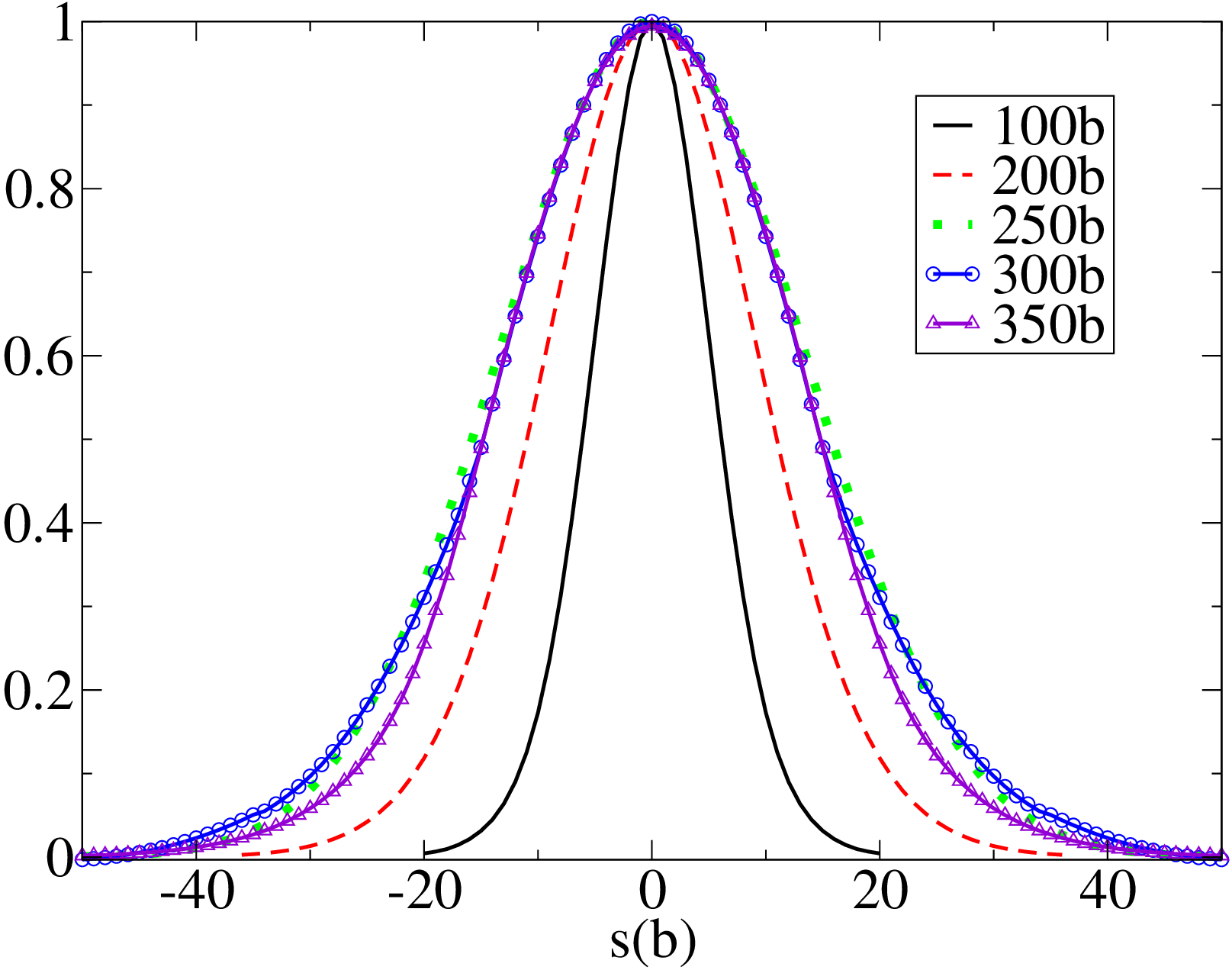}
\caption{\label{overlaps00} (color online): Comparison between the overlaps
$N_{00}(\bar{q}+s/2,\bar{q}-s/2)$ calculated for different $\bar{q}$
as functions of $s$ for $20\,\textrm{b}\le\bar{q}\le100\,\textrm{b}$
(left panel) and $100\,\textrm{b}\le\bar{q}\le350\,\textrm{b}$ (right
panel).}

\par\end{centering}

\centering{}%
\end{figure}

In Fig. \ref{overlaps00} the overlaps $N_{00}(\bar{q}+s/2,\bar{q}-s/2)$
between HFB solutions at different deformations are plotted as functions
of $s$ for different values of $\bar{q}$.

\begin{figure}[!h]

\begin{centering}
\includegraphics[width=7cm]{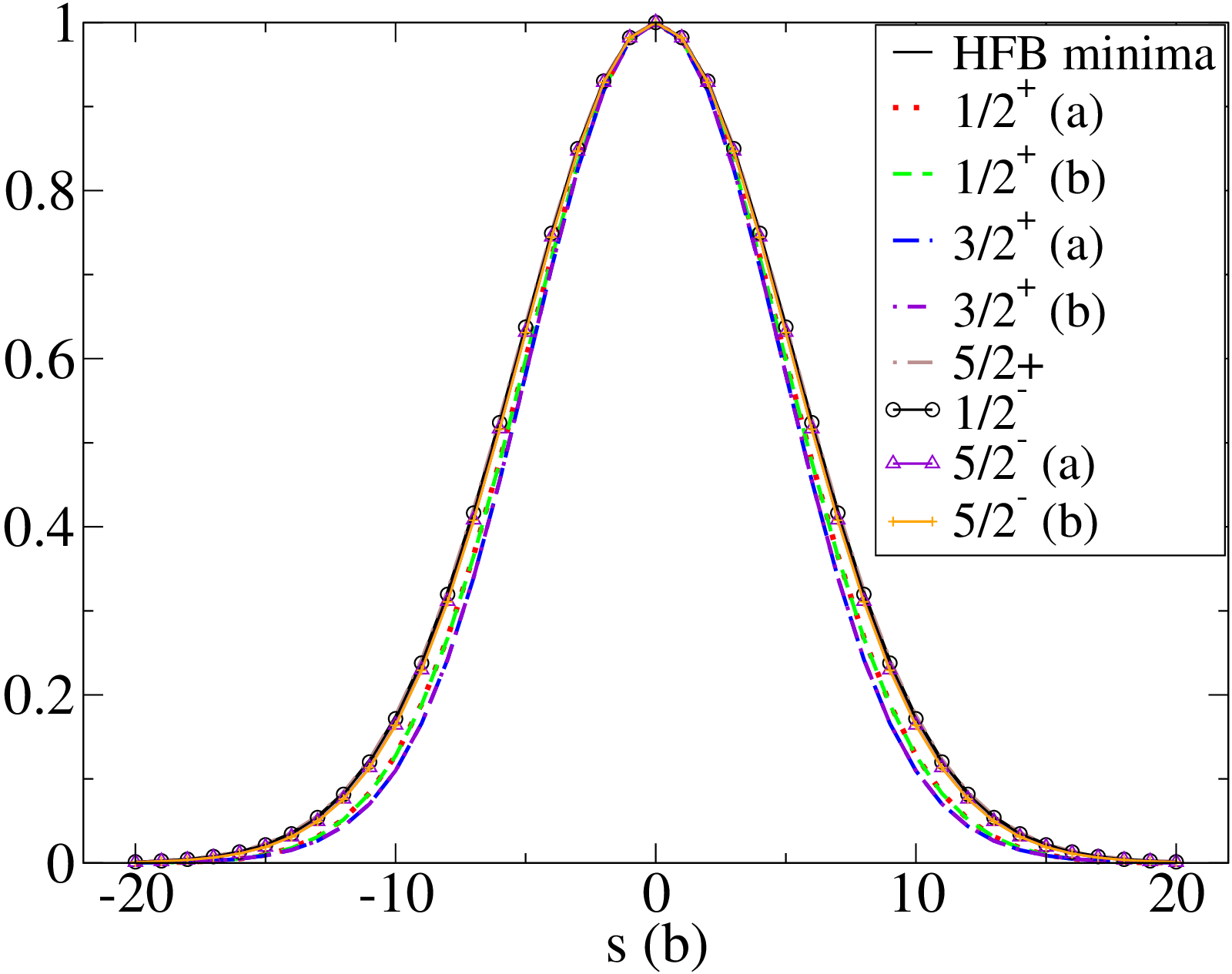}
\includegraphics[width=7cm]{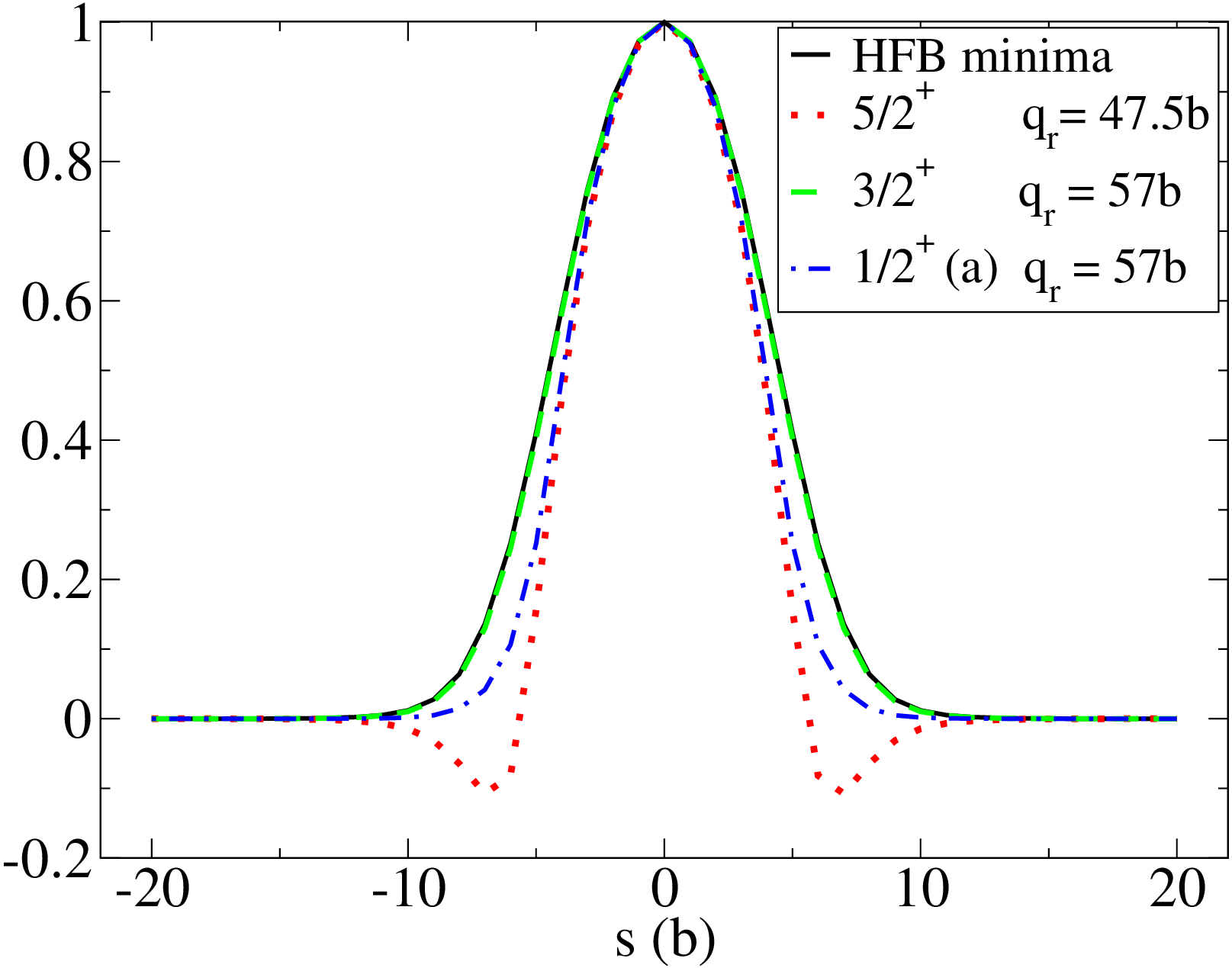}
\caption{\label{Nii} (color online): (Left panel) Overlap kernel $N_{ii}(\bar{q}+s/2,\bar{q}-s/2)$
for HFB solutions and different 2-qp excitations not involved in repulsions
(labeled by $K^{\pi}(i))$ at $\bar{q}=60\,\textrm{b}$, plotted as
functions of $s$. (Right panel) Overlap kernel $N_{ii}(\bar{q}+s/2,\bar{q}-s/2)$
for HFB solutions and different 2-qp excitations involved in repulsions
(labeled by $K^{\pi}(i)$, and $q_{r}$ the deformation where the
repulsion occurs) at $\bar{q}=50\,\textrm{b}$, plotted as functions
of $s$. Letters are used to differentiate same $K^{\pi}$ excitations
in both figures.}

\par\end{centering}

\centering{}%
\end{figure}

At first glance, all the overlaps shown in Fig. \ref{overlaps00}
suggest that they are close to a Gaussian shape, independently of
the mean quadrupole deformation $\bar{q}$. Although the SCIM formalism
derived in this paper does not require that the overlaps be Gaussian,
it is instructive to check the validity of this assumption, as a further
motivation for the more general moment-based approach we use. The
width of the overlap is found to strongly depend on the deformation.
In general, the width increases with the deformation, with local minima
all along the barrier. The extent to which the Gaussian shape is a
good approximation can be investigated by comparing different ways
to associate a width to these overlaps. In fact, assuming that they
can be represented by a Gaussian \begin{equation}
I(q,q')=e^{-(q-q')^{2}/2\sigma^{2}},\end{equation}
 their moments of even order should verify the relations \begin{eqnarray}
N^{(0)}(\bar{q}) & = & \sqrt{2\pi}\sigma(\bar{q})\nonumber \\
N^{(2n)}(\bar{q}) & = & (-1)^{n}\frac{(2n)!}{2^{n}n!}\sigma^{2n}(\bar{q})N^{(0)}(\bar{q})\label{widths}\end{eqnarray}
 which provide different ways to extract the width $\sigma(\bar{q})$.
The same procedure can be applied a priori to all the diagonal terms
of $N(\bar{q}+s/2,\bar{q}-s/2)$. In Fig. \ref{Nii} (left panel)
a few diagonal elements of the overlap matrix are plotted at $\bar{q}=50\,\textrm{b}$
where a comparison is made between overlaps of HFB vacuum and $2$
qp states. The different overlaps are very close together, but we
notice some difference in the tails. 
As we will see latter, these small deviations in the tails have a
significant influence on the value of the second-order moments. 
Furthermore, we find that level crossings do not generate any sizable
change in $N_{ii}^{(n)}(\bar{q})$, whereas repulsions are responsible
for a quite rapid but continuous change in the overlap. 
 Three different cases are plotted in Fig. \ref{Nii} (right panel).
First, we observe that the overlap (plotted as dashed line) of the
excitation built with two $K_{i}^{\pi}=3/2^{+}$ qp, one of them being
involved in a repulsion at $q_{r}=57\,\textrm{b}$, is not sizably
disturbed by the repulsion. Conversely, for the same configuration
-- one single excitation involved in a repulsion at $q_{r}=57\,\textrm{b}$
-- the overlap built with $K_{i}^{\pi}=1/2^{+}$ qp is sharper (smaller
width) than the previous one. 
 The most pathological case is found for the overlap built this time
with two $5/2^{+}$ excitations involved in a mutual repulsion at
$q_{r}=47.5\,\textrm{b}$, close to $\bar{q}=50\,\textrm{b}$. This
overlap even becomes negative in the region close to the level repulsion.
The comparison with a Gaussian overlap is therefore clearly unjustified
here. Let us note that these numerical results are in agreement with
the analysis of the repulsions we made in Section \ref{gene}. 

\begin{figure}[!h]

\begin{centering}
\includegraphics[width=7cm]{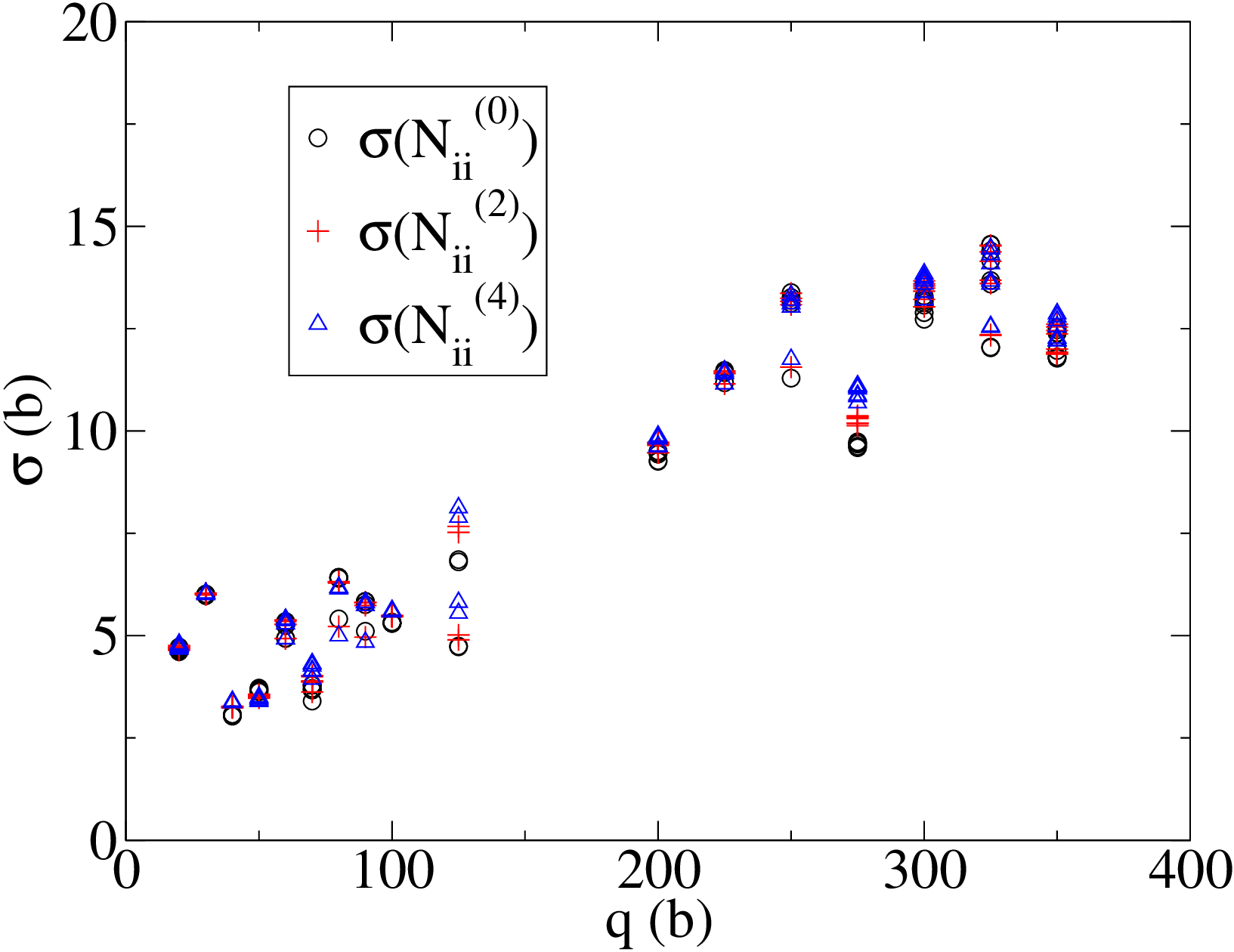}
\includegraphics[width=7cm]{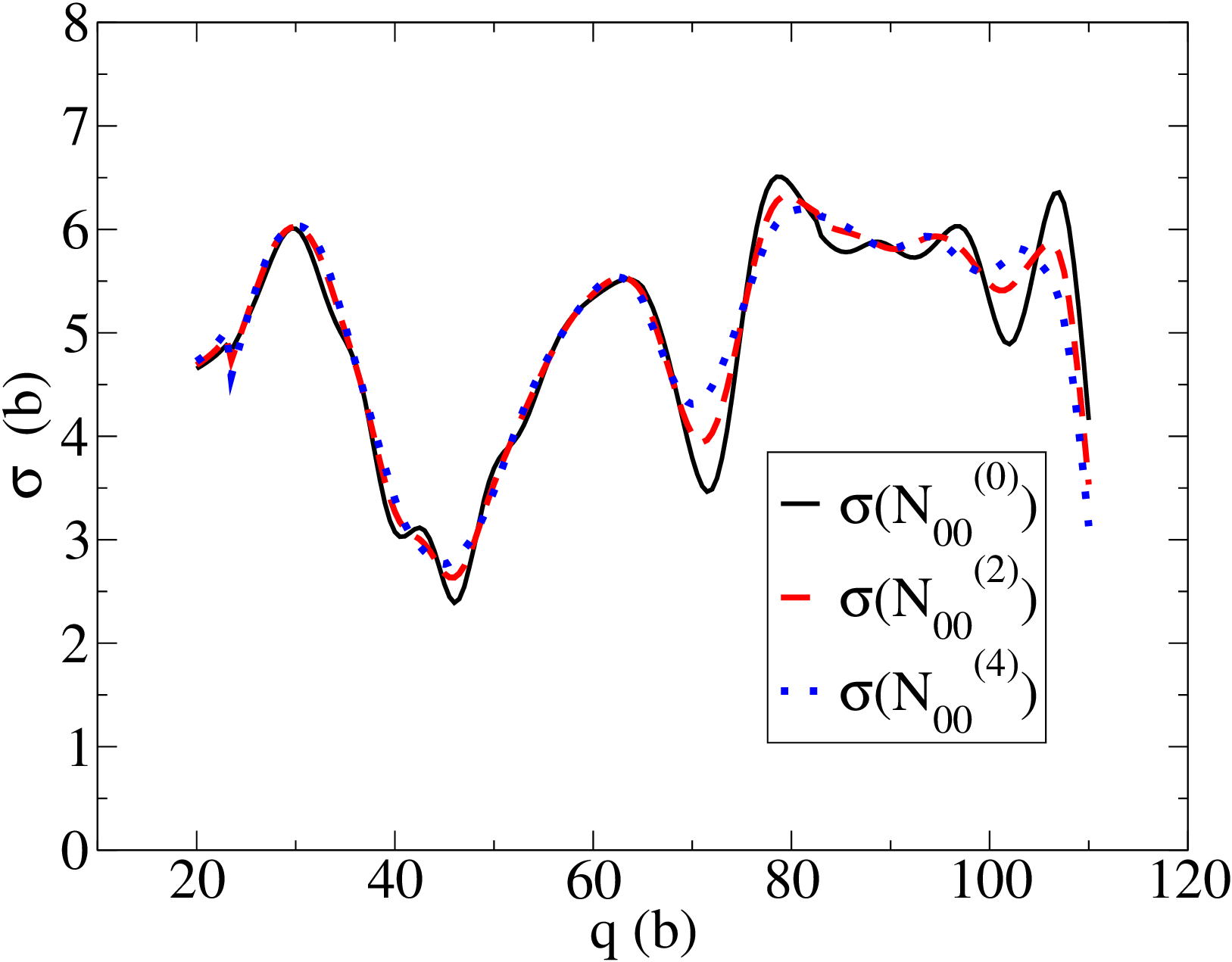} \caption{\label{width} (color online) Widths of the overlap extracted from
the moments of zero, second and fourth orders, as functions of the
elongation, in black circle, red cross and blue triangle, respectively.
Full lines correspond to the widths from $N_{00}^{(n)}$ terms and
circles to diagonal terms $N_{ii}^{(n)}$, $i\ne0$. Left panel: Results
for a large range of deformation $[0,350\,\textrm{b}]$. Right panel:
Widths from $N_{00}^{(n)}$ in the region $[0,110\,\textrm{b}]$. }

\par\end{centering}

\centering{}%
\end{figure}

The widths, obtained from Eq. (\ref{widths}) using the moments of
zero, second and fourth orders are plotted in Fig. \ref{width} as
functions of the deformation for the overlaps $N_{00}(\bar{q}+s/2,\bar{q}-s/2)$
and $N_{ii}(\bar{q}+s/2,\bar{q}-s/2)$. %

Particular attention has been paid to the widths extracted from the
$N_{00}^{(n)}(q)$ overlaps: calculations of the widths have been
performed in the range $[20\,\textrm{b},110\,\textrm{b}]$ with a
step of $0.5\,\textrm{b}$. On the right panel of Fig. \ref{width}
we observe that the three widths for $N_{00}^{(n)}(q)$ do not differ
too much in most cases, which corroborates the fact that the overlaps
are reasonably close to those from a Gaussian shape. For the lowest
deformations, the largest differences occur when the width is minimum,
at $46\,\textrm{b}$, $72\,\textrm{b}$ and $102\,\textrm{b}$ with
a maximal deviation of $22\%$ at $72\,\textrm{b}$, between the widths
extracted from the moments of zero and fourth orders.

Widths of the overlaps $N_{ii}(\bar{q})$, with $i\ne0$ are also
plotted up to $350\,\textrm{b}$ on the left panel of Fig. \ref{width}.
Since the excitations involved in repulsions do not lead to Gaussian
shapes, as illustrated in Fig. \ref{Nii}b), they are not taken into
account in this analysis. The maximal deviation between the widths
obtained from $N^{(0)}(\bar{q})$, $N^{(2)}(\bar{q})$ and $N^{(4)}(\bar{q})$
for a given excitation is found to be about $19\%$. The average of
this maximal deviation over all the different excitations and over
all the deformations is about $5\%$. However, we emphasize that these
deviations will pose no problem in our approach since we do not rely
on a Gaussian-overlap assumption at all.

\subsubsection{Calculation of the non-diagonal terms $N_{ij}(\bar{q}+s/2,\bar{q}-s/2)$
\label{calcNij}}

\begin{figure}[!h]

\begin{centering}
\includegraphics[width=7cm]{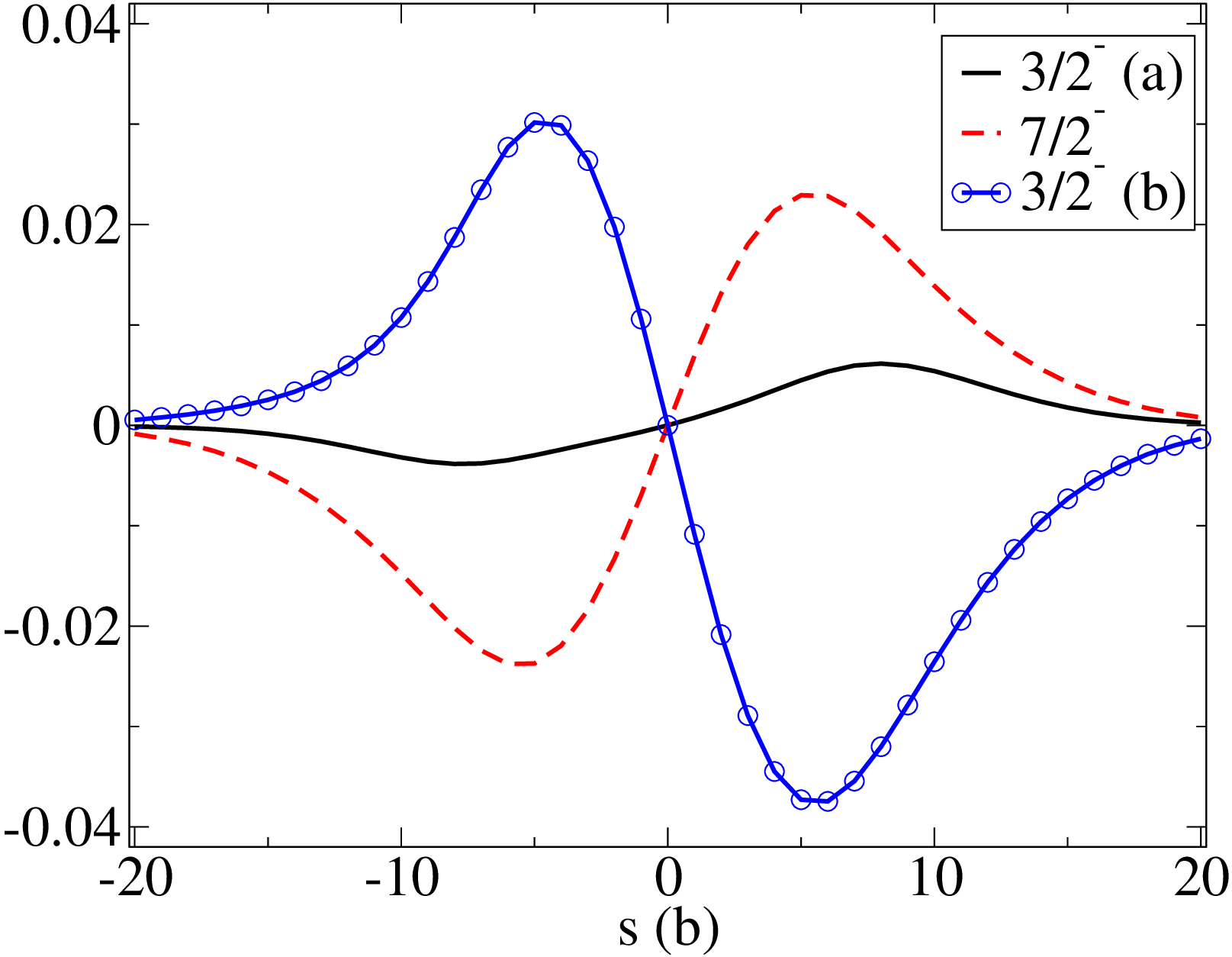}
\includegraphics[width=7cm]{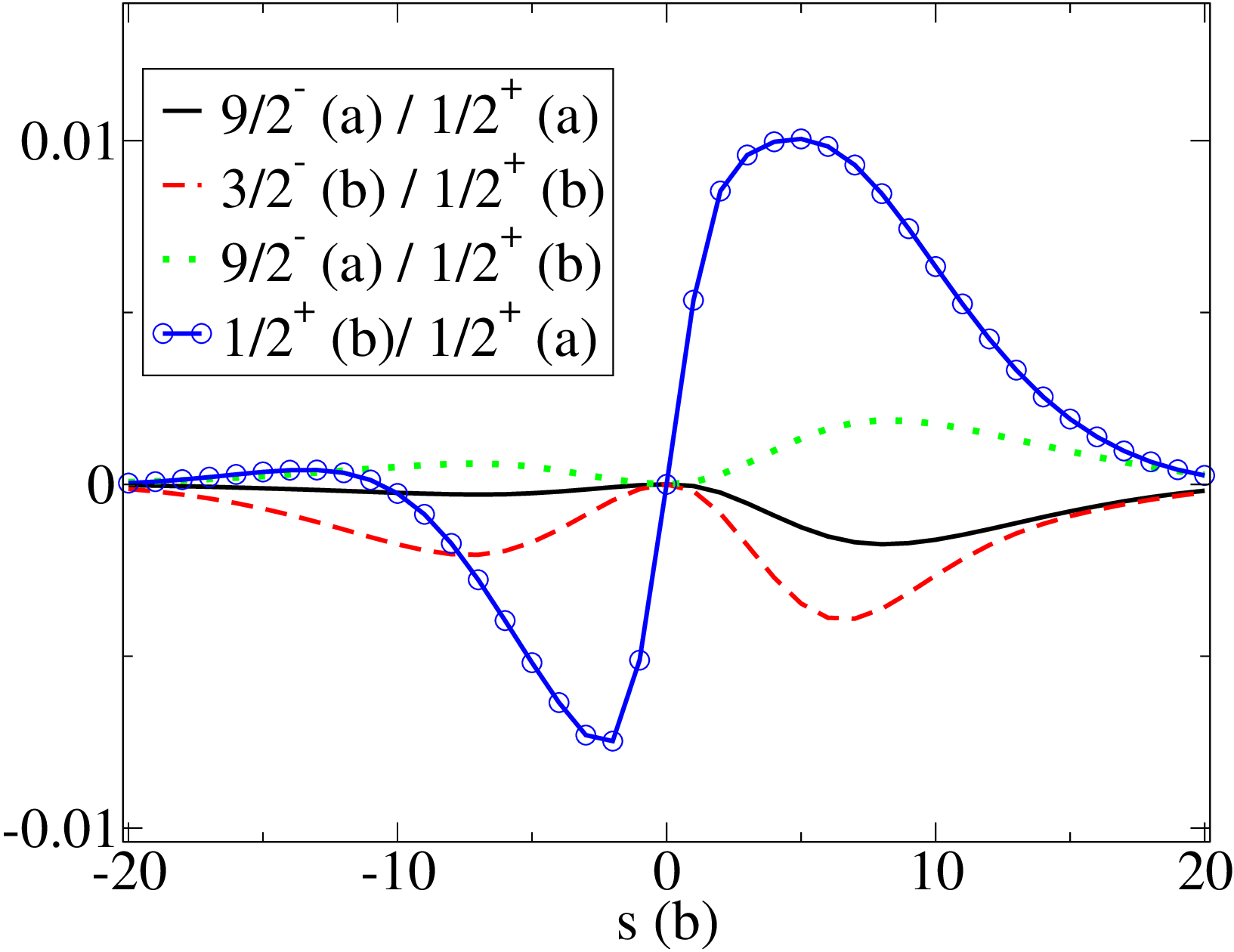}
\caption{\label{Nij} (Color online) Overlaps $N_{0i}(\bar{q}+s/2,\bar{q}-s/2)$
(left panel) and $N_{ij}(\bar{q}+s/2,\bar{q}-s/2)$ (right panel)
at $\bar{q}=30b$, for different excitations labeled by $K^{\pi}$
as a function of $s$. Letters are used to differentiate same $K^{\pi}$
excitations. }

\par\end{centering}

\centering{}%
\end{figure}

Non diagonal overlaps $N_{ij}(\bar{q}+s/2,\bar{q}-s/2)$ ($i\ne j$)
are plotted in Fig. \ref{Nij}. $N_{0i}(\bar{q}+s/2,\bar{q}-s/2)$
are displayed in the left panel, and $N_{ij}(\bar{q}+s/2,\bar{q}-s/2)$
in the right panel. For the sake of visibility, only a few overlaps
are drawn.

At this point let us emphasize that non-diagonal overlaps have no
a-priori reason to be even or odd functions of $s$. However, they
must be zero for $s=0$, since, at the same deformation, 2-qp excitations
are orthogonal to the ground state and any other 2-qp excitation.
We observe on this plot that in fact they are odd or even functions
of $s$ near the origin, depending on the indices $i$,$j$ of the
excitations. Notice that the amplitudes of these non-diagonal overlaps
are much smaller than the diagonal ones. Furthermore, the overlaps
between two excited states are also small in comparison to those between
the ground state and excited states. The exceptions are found for
qp states having the same quantum numbers $K^{\pi}(i)=K^{\pi}(j)$,
as shown by the blue curve with open circles in Fig. \ref{Nij}b).
This result is not surprising since, using the generalized Wick theorem
when $K^{\pi}(i)\ne K^{\pi}(j)$, the following relation is found:
\begin{eqnarray}
N_{ij}(\bar{q}+s/2,\bar{q}-s/2)=\frac{1}{N_{00}(\bar{q}+s/2,\bar{q}-s/2)}\nonumber \\
\times N_{0i}(\bar{q}-s/2,\bar{q}+s/2)N_{0j}(\bar{q}+s/2,\bar{q}-s/2).\end{eqnarray}
 We also mention here that the overlap matrix is not block-diagonal
in isospin, since \[
\mathcal{N}_{ij}^{\tau_{i}\tau_{j}}(q,q')=\bra{\Phi(q)}|\eta_{i_{2}}^{\tau_{i}}\eta_{i_{1}}^{\tau_{i}}\eta_{j_{1}}^{\tau_{j}+}\eta_{j_{2}}^{\tau_{j}+}\ket{\Phi(q')}\]
 is non zero for $\tau_{i}\ne\tau_{j}$.

\begin{figure}[!h]

\begin{centering}
\includegraphics[width=8cm]{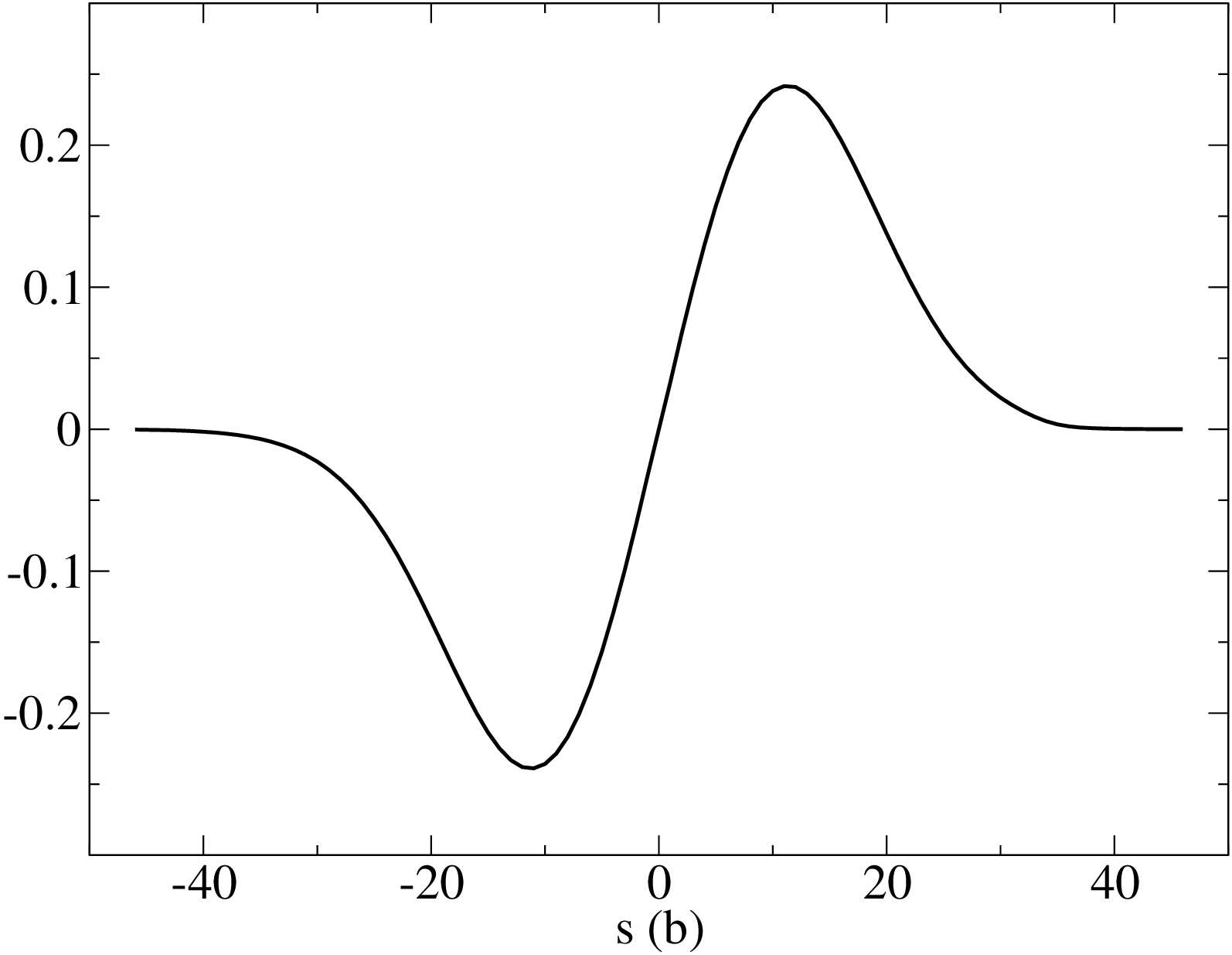} 
\caption{\label{rep224} Non diagonal term $N_{ij}(\bar{q})$ built with excitations
involved in a repulsion at $q_{r}=218\,\textrm{b}$.}

\par\end{centering}

\centering{}%
\end{figure}

Conclusions obtained for the diagonal terms of $N^{(0)}(\bar{q})$
about repulsions and crossings still apply here: non-diagonal overlaps
are not disturbed by crossing levels but can vary rapidly when a repulsion
occurs. In Fig. \ref{rep224} the overlap $N_{ij}$ for a couple of
2-qp excitations $i=(i_{1},i_{2})$ and $j=(j_{1},j_{2})$ with $K_{i(j)}^{\pi}=1/2^{-}$
is depicted. In this case, $i_{1}$ and $i_{2}$ are both involved
in a mutual repulsion at $q_{r}=218\,\textrm{b}$ and $i_{2}=j_{2}$.
The amplitude of this overlap is much larger than that of the standard
(without repulsion) non diagonal overlap. Let us mention that this
kind of configuration is encountered several times along the deformation
and corroborates our observation in Section \ref{sel2qp} that the
qp states $i_{1}$ and $j_{1}$ are mixed in the repulsion area.

\subsubsection{Numerical study of the Moments $N_{ij}^{(n)}(\bar{q})$, $n=0,1,2$
\label{anamom}}

According to our approximation (see Eq.(\ref{calcJ2})), the overlap
kernel has to be analyzed through its moments up to the second order.
At this point it is useful to introduce some important properties
of the moments defined in Section \ref{SME}. Their demonstration
being straightforward we only list them here:\\
 - All moments of a Hermitian operator are Hermitian: \[
\hat{A}^{+}=\hat{A}\Rightarrow A^{(p)+}=A^{(p)}.\]
 By applying this property to the overlap kernel, whose matrix elements
are real for reasons given at the beginning of this section, the matrices
associated with even-order moments are found to be real symmetric
while those associated with odd-order moments are imaginary antisymmetric.\\
 - Finally, let us indicate that the even moments of an operator
have the same parity with respect to the time reversal symmetry ($\hat{T}$)
as the operator considered, while the opposite is true for odd moments
\[
\hat{T}^{+}\hat{A}\hat{T}=\epsilon\hat{A}\Rightarrow\hat{T}^{+}A^{(p)}\hat{T}=(-1)^{p}\epsilon A^{(p)},\]
where $\epsilon=\pm1$.

With these properties established, we can now analyze numerical calculations
of each of the three moments $N_{ij}^{(n)}(\bar{q})$ for $n=0,1,2$
.

\paragraph{The Zero order moment matrix $N^{(0)}(\bar{q})$}


This matrix plays an essential role in the formalism as observed in
Eqs. (\ref{calcJ2}) and (\ref{eq:EqSch2-ord2}). In fact it is the
inverse of its square root which occurs in these expressions and whose
determination requires in principle the diagonalization of $N^{(0)}(\bar{q})$
with a matrix $U(\bar{q})$ which depends itself on the deformation
$\bar{q}$. As a consequence, the transformation of the symmetric
operators in Eqs.(\ref{calcJ2}) and (\ref{eq:EqSch2-ord2}) to the
new representation necessitates the calculation of the first and second
derivatives of $U(\bar{q})$ thereby including a number of terms which
significantly increase the size of the expressions (\ref{calcJ2})
and (\ref{eq:EqSch2-ord2}). In light of the numerical calculations
presented above, we propose instead a set of reasonable approximations
to avoid these unnecessary complications.

\begin{figure}[!h]

\begin{centering}
\includegraphics[width=7cm]{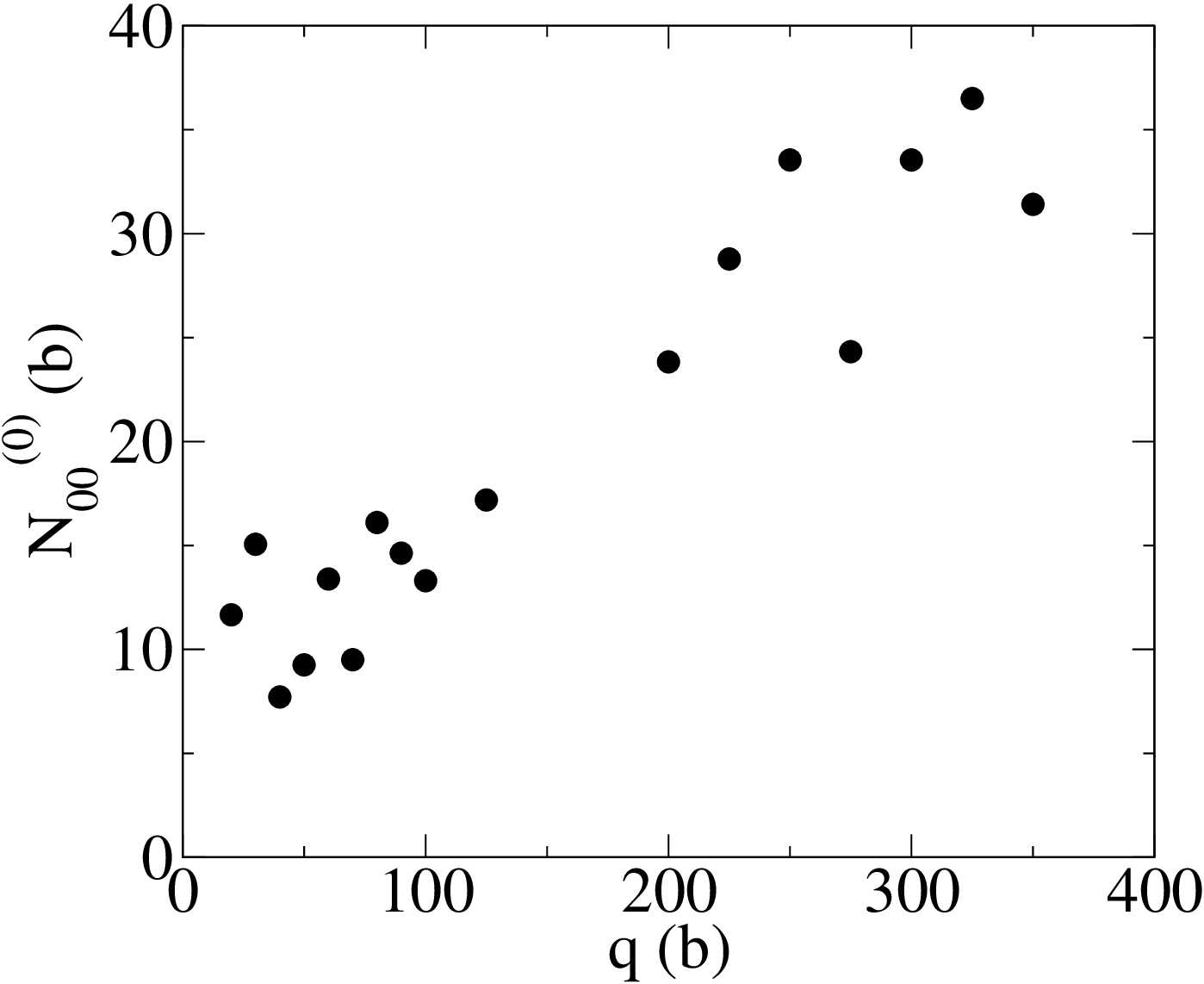}
\includegraphics[width=8.2cm]{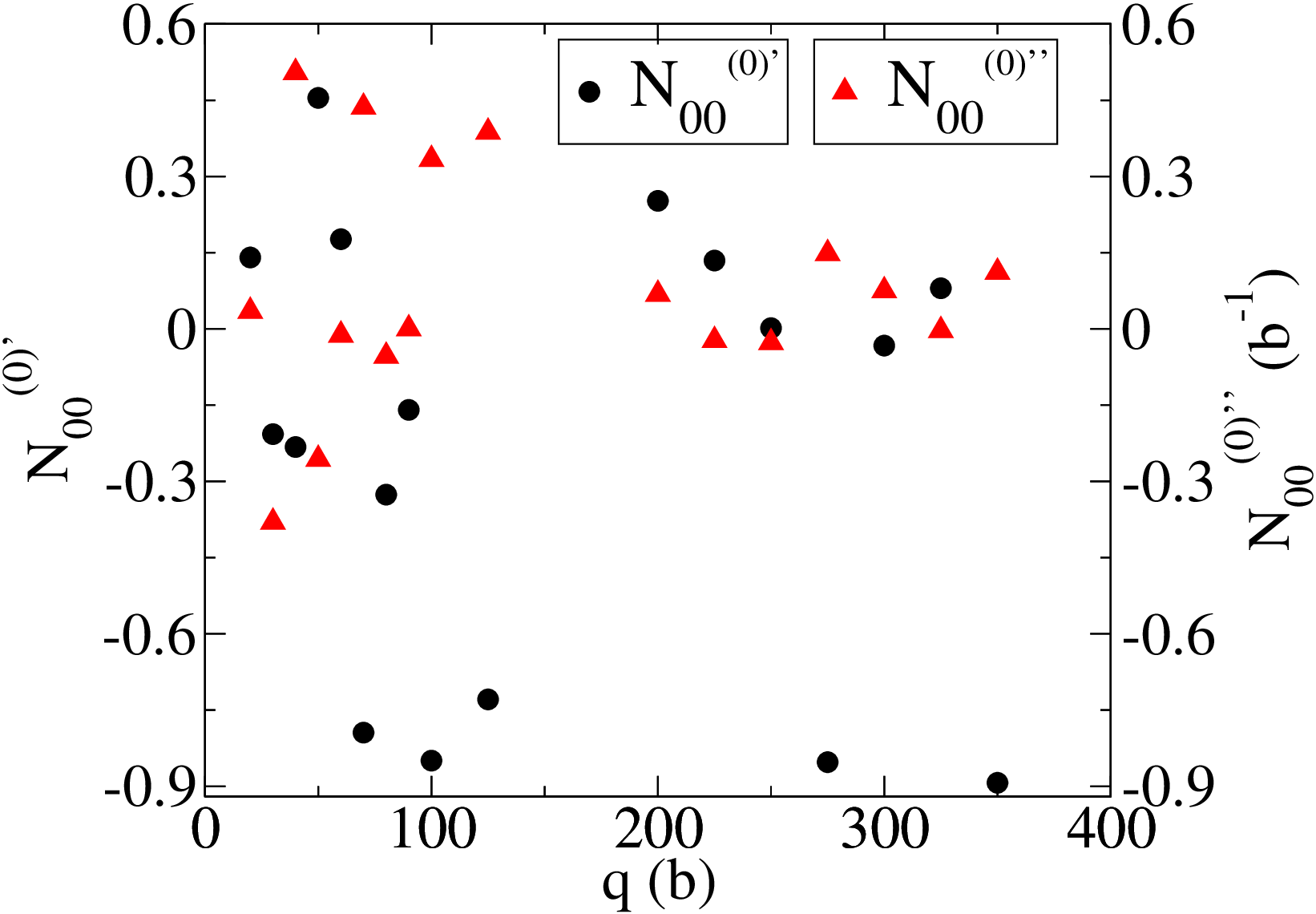}
\caption{\label{mom0_00} (left panel) $N_{00}^{(0)}(\bar{q})$ and its first
and second derivatives (right panel) as functions of the deformation. }

\par\end{centering}

\centering{}%
\end{figure}

First, in Fig. \ref{mom0_00}, the moment $N_{00}^{(0)}(\bar{q})$,
as well as its first and second derivatives, all of which intervene
in various quantities, are shown at different deformations.

The moment generally increases between $8\,\textrm{b}$ and $37\,\textrm{b}$
as a function of the quadrupole deformation. Its first two derivatives
$N_{00}^{(0)'}(\bar{q})$ and $N_{00}^{(0)''}(\bar{q})$ are found
to be much smaller by at least one order of magnitude. Their values
vary rapidly in the interval $[-0.9,0.5]$ for $N_{00}^{(0)'}(\bar{q})$
and $[-0.4\,\textrm{b}^{-1},0.6\,\textrm{b}^{-1}]$ for $N_{00}^{(0)''}(\bar{q})$.

\begin{table*}
\begin{tabular}{|l|r|r|r|r|r|r|r|r|r|}
\hline 
$K^{\pi}$  &  $0$  & $5/2^- (a)$ & $5/2^- (b)$ &$1/2^-$ & $1/2^+ (a)$&$1/2^+ (b)$ &\begin{tabular}{r r}$3/2^+ (a)$ \\$q_r=57b$ \end{tabular} &\begin{tabular}{r r}$3/2^+ (b)$ \\$q_r=57b$ \end{tabular}& $5/2^+ $\tabularnewline
\hline 
$0$ & 13 &  &  &  &  &  &  &  & \tabularnewline
\hline 
$5/2^{-}(a)$ &$  2.1\times 10^{-2} $& 13 &  &  &  &  &  &  & \tabularnewline
\hline 
$5/2^{-}(b)$ &$ -1.7\times 10^{-2} $&$ -6.0\times 10^{-3} $& 13 &  &  &  &  &  & \tabularnewline
\hline 
$1/2^{-}   $ &$ -1.2\times 10^{-2} $&$  4.3\times 10^{-3} $&$  5.4\times 10^{-3} $& 13 &  &  &  &  & \tabularnewline
\hline 
$1/2^{+}(a)$ &$  6.8\times 10^{-2} $&$ -2.1\times 10^{-3} $&$ -3.3\times 10^{-3} $&$  3.5\times 10^{-3}  $& 12 &  &  &  & \tabularnewline
\hline 
$1/2^{+}(b)$ &$ -7.4\times 10^{-2} $&$  8.1\times 10^{-3} $&$  1.1\times 10^{-2} $&$ -1.3\times 10^{-2}  $&$ -2.2\times 10^{-3} $& 12 &  &  & \tabularnewline
\hline 
$3/2^{+}(a), q_r=57b$ &$ -4.5\times 10^{-2} $&$  6.3\times 10^{-3} $&$  8.4\times 10^{-3} $&$ -1.0\times 10^{-2} $&$  4.5\times 10^{-3} $&$ -2.0\times 10^{-2} $& 12 &  & \tabularnewline
\hline 
$3/2^{+}(b), q_r=57b$ &$ -1.3\times 10^{-1} $&$ -3.1\times 10^{-3} $&$ -2.6\times 10^{-3} $&$  4.2\times 10^{-3} $&$ -4.0\times 10^{-3} $&$ 9.2\times 10^{-3}  $&$ 5.2\times 10^{-3} $& 12 & \tabularnewline
\hline 
$5/2^{+}   $ &$  8.9\times 10^{-3} $&$  5.4\times 10^{-3} $&$  6.6\times 10^{-3} $&$ -8.3\times 10^{-3} $&$ 4.6\times 10^{-3} $&$ -1.7\times 10^{-2} $&$ -1.3\times 10^{-2} $&$ 4.5\times 10^{-3} $& 13 \tabularnewline
\hline

\end{tabular}\captionof{table}{Matrix elements of $N^{(0)}(\bar{q})$ at $\bar{q}=60\,\textrm{b}$.
Centers of repulsion $q_{r}$ are mentioned if needed. Letters are
used to differentiate same $K^{\pi}$ excitations. }\label{N0} 
\end{table*}

In Table \ref{N0} we give the matrix elements of $N^{(0)}(\bar{q})$
for different excitations at $\bar{q}=60\,\textrm{b}$. At this deformation,
we observe that the diagonal elements do not depend too much on the
0 or 2 qp states under consideration: they are found to be between
$12.0\,\textrm{b}$ to $13.4\,\textrm{b}$. Such a feature is also
observed for most of the deformations. However, let us mention that
in a few cases, repulsions between single-particle states induce larger
variations of the diagonal elements $N_{ii}^{(0)}(\bar{q})$. This
observation suggests we should introduce the diagonal terms $N_{AV}^{(0)}(\bar{q})$
defined as \[
N_{AV}^{(0)}(\bar{q})=\frac{1}{N}Tr(N^{(0)}(\bar{q}))\]
 with $N$ the dimension of the matrix $N^{(0)}(\bar{q})$. With such
a definition, the matrix $N^{(0)}(\bar{q})$ can then be written in
the form \[
N^{(0)}(\bar{q})=N_{AV}^{(0)}(\bar{q})[I+\Delta(\bar{q})],\,\,\,\Delta(\bar{q})=\frac{N^{(0)}(\bar{q})-N_{AV}^{(0)}(\bar{q})I}{N_{AV}^{(0)}(\bar{q})}\]
 The diagonal and off-diagonal matrix elements of $\Delta$ are all
found to be small for all deformations , and therefore the inverse
of $N^{(0)}(\bar{q})$ can be calculated by means of the series expansion
of $(I+\Delta(q))^{-1/2}$. 
 As a result of this study, it appears that considering this inverse
square root as diagonal is a reasonable approximation.

At this point, we introduce normalized moments for non zero order
moments of the overlap kernel since they are the quantities of interest
in the formalism developed in sections \ref{deterJ} and \ref{SE},
\begin{equation}
\bar{N}^{(p)}(\bar{q})=\frac{1}{\sqrt{N^{(0)}(\bar{q})}}N^{(p)}(\bar{q})\frac{1}{\sqrt{N^{(0)}(\bar{q})}},\,\, p\ge1.\label{defNbar}\end{equation}
 Taking advantage of our previous discussion, these moments are calculated
in the approximation that the inverse square root of $N^{(0)}(\bar{q})$
is diagonal. Their properties are discussed in the next paragraph.

\paragraph{First order matrix moment $\bar{N}^{(1)}(\bar{q})$}

The matrix elements of normalized first order moment $\bar{N}^{(1)}(\bar{q})$
are given in Table \ref{N1} for different excitations at $\bar{q}=60b$.
Since $\bar{N}^{(1)}(\bar{q})$ is antisymmetric, the diagonal terms
are zero. From Eq. (\ref{defNbar}), it is clear that the value of
$\bar{N}_{ij}^{(1)}(\bar{q})$ depends on the amplitude of the overlap
$N_{ij}^{(1)}(\bar{q})$. As shown in Table \ref{N1}, the absolute
value of $\bar{N}_{ij}^{(1)}(\bar{q})$ is generally small although
some specific excitations associated with level repulsions lead to
non negligible values.

\begin{table*}
\begin{tabular}{|l|r|r|r|r|r|r|r|r|r|}
\hline 
$K^{\pi}$  &  $0$  & $5/2^- (a)$ & $5/2^- (b)$ &$1/2^-$ & $1/2^+ (a)$&$1/2^+ (b)$ & \begin{tabular}{r r}$3/2^+ (a)$ \\$q_r=57b$ \end{tabular} &  \begin{tabular}{r r}$3/2^+ (b)$ \\$q_r=57b$\end{tabular} & $5/2^+ $\tabularnewline
\hline 
$0$  & 0.0 &  &  &  &  &  &  &  & \tabularnewline
\hline 
$5/2^{-}(a)$ &$ -7.9\times 10^{-2}$  & 0.0 &  &  &  &  &  &  & \tabularnewline
\hline 
$5/2^{-}(b)$ &$ -9.8\times 10^{-2}$& $-9.9\times 10^{-1}$ & 0.0 &  &  &  &  &  & \tabularnewline
\hline 
$1/2^{-}$ & $1.2\times 10^{-1}$ & $3.7 \times 10^{-4}$& $-8.1 \times 10^{-4}$& 0.0 &  &  &  &  & \tabularnewline
\hline 
$1/2^{+}(a)$ & $-6.9\times 10^{-2}$ & $1.1 \times 10^{-3}$& $2.0\times 10^{-3}$ & $-1.8 \times 10^{-3}$& 0.0 &  &  &  & \tabularnewline
\hline 
$1/2^{+}(b)$ & $2.5\times 10^{-1}$ & $-5.9 \times 10^{-4}$& $-2.5 \times 10^{-3}$& $1.5\times 10^{-3}$ & $7.3 \times 10^{-2}$& 0.0 &  &  & \tabularnewline
\hline 
$3/2^{+}(a), q_r=57b$ & $1.9 \times 10^{-1}$& $-3.1\times 10^{-4}$ & $-1.9\times 10^{-3}$ & $-1.2\times 10^{-3}$ & $-1.8 \times 10^{-3}$& $4.0 \times 10^{-4}$& 0.0 &  & \tabularnewline
\hline 
$3/2^{+}(b), q_r=57b$ & $-7.3 \times 10^{-2}$& $-2.6\times 10^{-3}$ & $-2.2 \times 10^{-3}$& $-3.7 \times 10^{-3}$& $3.7\times 10^{-3}$ & $-7.6 \times 10^{-3}$& $2.5 \times 10^{+0}$& 0.0 & \tabularnewline
\hline 
$5/2^{+}$ & $1.5\times 10^{-1}$ & $-8.4 \times 10^{-4}$& $-4.5\times 10^{-4}$ & $6.1\times 10^{-4}$ & $-2.5 \times 10^{-3}$& $2.8 \times 10^{-3}$& $2.0\times 10^{-3}$ & $3.6\times 10^{-3}$ & 0.0 \tabularnewline
\hline
\end{tabular}\captionof{table}{Matrix elements of normalized first-order moment
$\bar{N}^{(1)}(\bar{q})$ for different excitations at $\bar{q}=60\,\textrm{b}$.
All the values are divided by i in order to be real. Centers of repulsion
$q_{r}$ are mentioned if needed. Letters are used to differentiate
same-$K^{\pi}$ excitations.}\label{N1} 
\end{table*}

Over the whole range of deformation, the most singular case concerns
two excitations involved in a mutual repulsion for which $|N_{ij}^{(1)}(\bar{q})|=6.9\,\textrm{b}$.
 All along the deformation, only $27$ matrix elements $|N_{ij}^{(1)}(\bar{q})|$
(out of $612$) are higher than $1\,\textrm{b}$. 

\paragraph{Second order moment $\bar{N}^{(2)}(\bar{q})$}

As we will see in Section (\ref{deterJ}), the derivation of the inverse
of $\hat{J}_{1/2}(\bar{q})=(I+\hat{u}(\bar{q}))^{1/2}$ requires the
determination of not only the second-order moments, but also its first
and second derivatives.

\begin{figure}[!h]

\begin{centering}
\includegraphics[width=7cm]{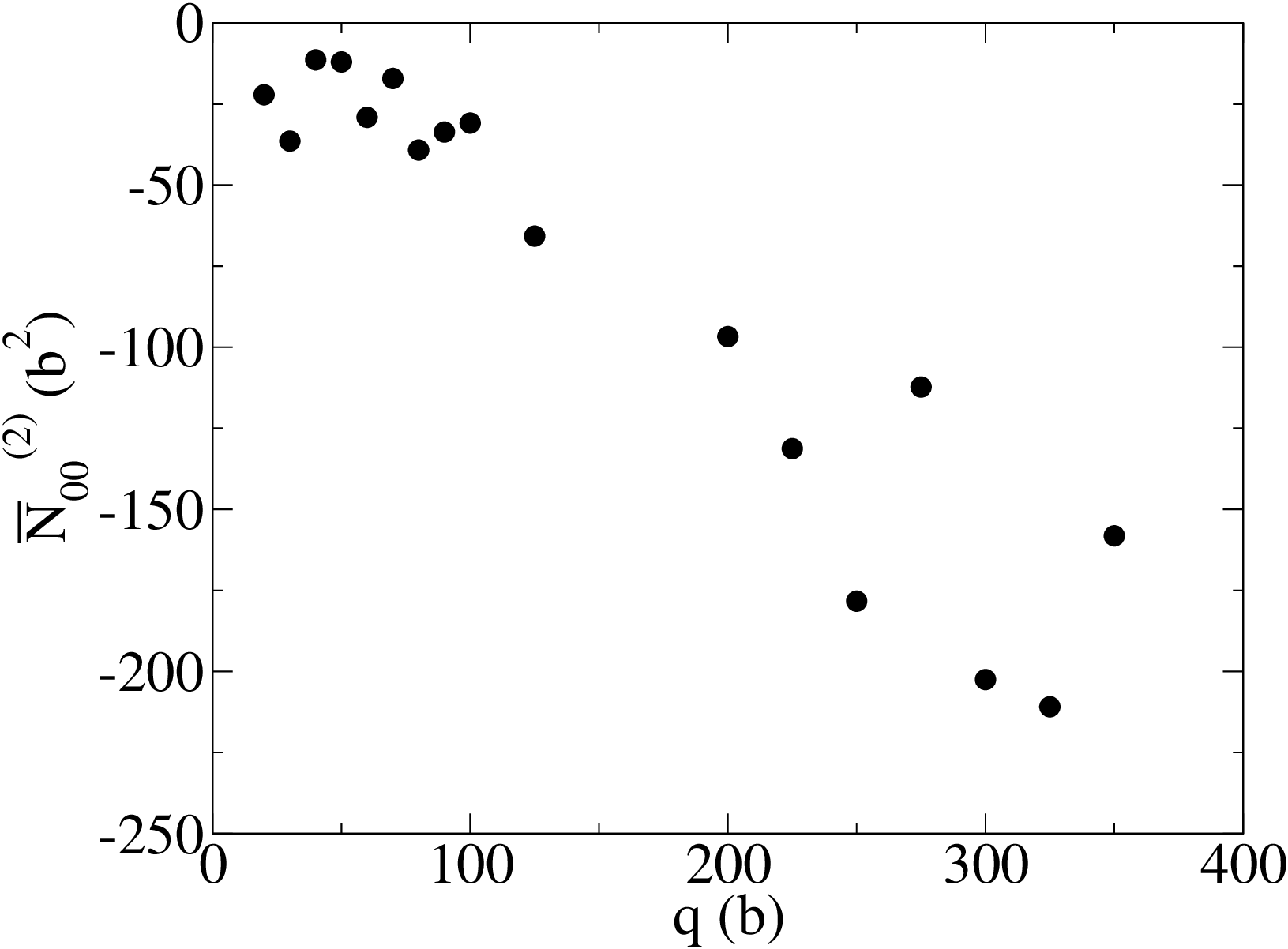}
\includegraphics[width=7cm]{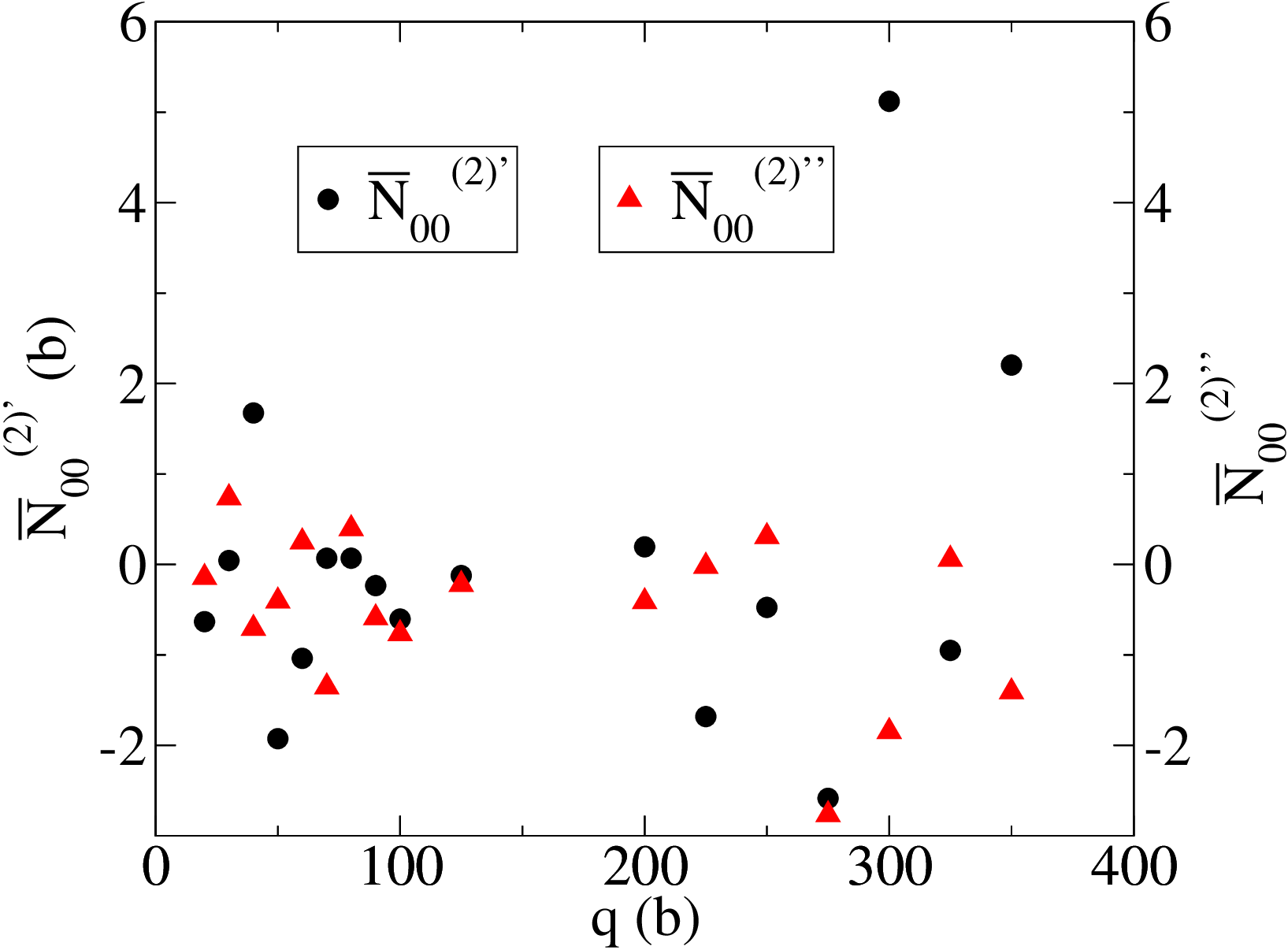}
\caption{ \label{mom2_00} (Color online) The normalized second order moment
$\bar{N}^{(2)}(\bar{q})$ (left panel) and its first (black filled
circles) and second (red filled triangles) derivatives (right panel)
of the non excited overlap along the deformation. }

\par\end{centering}

\centering{}%
\end{figure}

For the sake of brevity, we only display in Fig. \ref{mom2_00}a)
and \ref{mom2_00}b) those quantities in the case of the diagonal
element in the ground state.

In Fig. \ref{mom2_00}a) we notice that all these quantities, not
only vary rapidly with the deformation but they also undergo oscillations
of very large amplitude. The moment itself varies in between $-5\,\textrm{b}^{2}$
and $-211\,\textrm{b}^{2}$ in a range of deformations $[20\,\textrm{b},350\,\textrm{b}]$.
As for the first and second derivatives they are contained in the
range $[-3\,\textrm{b},6\,\textrm{b}]$ and $[-2.1,1]$ respectively
(Fig. \ref{mom2_00}b)).

In Table \ref{N2}, diagonal and non-diagonal normalized second-order
moments are given for the HFB minimum and the selected excitations
at $\bar{q}=80\,\textrm{b}$. Diagonal terms are negative due to our
definition of the second-order moment. As expected from our discussion
on non-diagonal matrix element of the overlap, most of the non-diagonal
terms are at least two orders of magnitude smaller than the diagonal
ones. It is also seen that they strongly depend on the excitations.
Very similar features are observed for all deformations up to $\bar{q}=350\,\textrm{b}$.
However, the maximum value of a non-diagonal term is found to be $\underset{i\neq j,\bar{q}}{\sup}|\bar{N}_{ij}^{(2)}(\bar{q})|=78.4\,\textrm{b}^{2}$
at $\bar{q}=250\,\textrm{b}$, which corresponds to $44\%$ of $\bar{N}_{00}^{(2)}(\bar{q})$
in this region. This case corresponds to $i=(i_{1},i_{2})$ and $j=(j_{1},j_{2})$
excitations for which repulsion occurs between $i_{1}$ and $j_{1}$
at $q_{r}=242\,\textrm{b}$ and between $i_{2}$ and $j_{2}$ at $q_{r}=265\,\textrm{b}$.
\begin{table*}
\begin{tabular}{|l|r|r|r|r|r|r|r|r|r|}
\hline 
$K^{\pi}$  &  $0$  & $3/2^- (a)$ & \begin{tabular}{r r}$3/2^- (b)$ \\$q_r=82.5b$ \end{tabular} &$1/2^-$ & \begin{tabular}{r r}$1/2^+ (a)$ \\$q_r=85b$ \end{tabular}&\begin{tabular}{r r}$1/2^+ (b)$ \\$q_r=85b$ \end{tabular} &\begin{tabular}{r r r}$1/2^+ (c)$ \\$q_r=85b$\\$q_r'=86b$ \end{tabular} &$5/2^+$ & $9/2^+ $\tabularnewline
\hline 
$0$ & -39 &  &  &  &  &  &  &  & \tabularnewline
\hline 
$3/2^{-}(a)$ & $1.1\times 10^{+0}$ & $-26$ &  &  &  &  &  &  & \tabularnewline
\hline 
$3/2^{-}(b)$ & $-1.9\times 10^{-1}$ & $1.2\times 10^{+0}$ & $-30$ &  &  &  &  &  & \tabularnewline
\hline 
$1/2^{-}   $ & $2.4\times 10^{-2}$ & $4.3\times 10^{-1}$ & $3.2\times 10^{-2}$ & $-39$ &  &  &  &  & \tabularnewline
\hline 
$1/2^{+}(a)$, $q_r=85b$ & $-3.7\times 10^{-1}$ & $1.1\times 10^{-1}$ & $2.7\times 10^{-3}$ & $6.3\times 10^{-3}$ & $-32$ &  &  &  & \tabularnewline
\hline 
$1/2^{+}(b)$, $q_r=85b$ & $-8.7\times 10^{-2}$ & $-2.7\times 10^{-1}$ & $-2.1\times 10^{-2}$ & $-1.9\times 10^{-2}$ & $-2.2\times 10^{-2}$ & $-32$ &  &  & \tabularnewline
\hline 
$1/2^{+}(c)$, \begin{tabular}{r r}$q_r=85b$\\$q_r'=86b$\end{tabular} & $1.1\times 10^{-1}$ & $-2.4\times 10^{-1}$ & $-1.6\times 10^{-2}$ & $-1.9\times 10^{-2}$ & $2.4\times 10^{0}$ & $2.3\times 10^{-1}$ & $-29$ &  & \tabularnewline
\hline 
$5/2^{+}   $ & $2.3\times 10^{-1}$ & $8.6\times 10^{-1}$ & $6.6\times 10^{-2}$ & $7.1\times 10^{-2}$ & $1.5\times 10^{-2}$ & $-4.5\times 10^{-2}$ & $-3.8\times 10^{-2}$ & $-39$ & \tabularnewline
\hline 
$9/2^{+}   $ & $2.3\times 10^{-1}$ & $2.9\times 10^{-1}$ & $2.1\times 10^{-1}$ & $2.5\times 10^{-2}$ & $8.5\times 10^{-3}$ & $-1.5\times 10^{-2}$ & $-1.4\times 10^{-2}$ & $5.7\times 10^{-2}$ & $-39$ \tabularnewline
\hline
\end{tabular}\captionof{table}{Matrix elements of normalized second order moment
$\bar{N}_{ij}^{(2)}(\bar{q})$ for different excitations at $\bar{q}=80\,\textrm{b}$.
Centers of repulsion $q_{r}$ are mentioned if needed. Letters are
used to differentiate same $K^{\pi}$ excitations.}\label{N2} 
\end{table*}

It is worth stressing that perturbations in the overlap matrix due
to repulsions, such as the change in amplitude of a non-diagonal term
as depicted in Fig. \ref{rep224}, or the reduction of the diagonal
terms (see Fig. \ref{Nii}b), disturbs the values of the moments,
which are the quantities of interest in the present formalism, in
a brutal but still continuous way .

\section{Study and determination of $\hat{J}_{\pm1/2}(\bar{q})$ \label{deterJ}}

In the previous section we have seen that the moments and their derivatives
vary rapidly as function of the deformation. As we will see in the
following these quantities occur everywhere in the calculation of
$\hat{J}_{\pm1/2}(\bar{q})$ (with $\hat{J}_{\pm1/2}(\bar{q})=\hat{J}^{\pm1/2}(\bar{q})$),
which seriously complicates the derivation of the inverse of the overlap
kernel. This question of the dependence of the moments on the deformation
has been studied in detail in the GOA \cite{Reinhard87}, and in the
symmetric-moment expansion for the one-dimensional case of no intrinsic
excitations \cite{Bauhoff80} (although, even in the one-dimensional
case, our solution below is more complete), but in the context of
the present study we have not found any information on this subject
in the open literature. In the discussion that follows, we will call
$[.P]^{(n)}$ a symmetric operator of order $n$ despite the fact
that, strictly speaking, an order cannot be attributed to such an
operator since it contains all orders up to $n$.

\subsection{Discussion and generalities about the inversion of the one-dimensional
overlap kernel}

The difficulty mentioned above already occurs in the standard (one-dimensional)
approach and consequently the discussion is first restricted to this
simple case. Let us recall here that when the moments are independent
of the deformation, the derivation of the operators in question is
straightforward. They are given by \cite{RingSchuck80} \[
\hat{J}_{\pm1/2}(\bar{q})=[1+\hat{u}(\bar{q})]^{\pm1/2}\simeq1\pm\frac{1}{2}\hat{u}(\bar{q})\]
 with \[
\hat{u}(\bar{q})=\frac{N^{(2)}(\bar{q})}{2N^{(0)}(\bar{q})}P^{2}.\]
 This form is consistent with an expansion of the kernel overlap up
to second order in the collective momentum $P$. Besides, we are justified
in stopping the expansion of $\hat{J}_{-1/2}(\bar{q})$ at the second
order in $P$ since, in the usual derivation of a Schrödinger equation,
one neglects the derivatives of the Hamiltonian kernel of order greater
than two. The situation is quite different here since symmetric operators
of order four give contributions to the expansion of the Hamiltonian
kernel involving derivatives less or equal than two. In order to go
beyond this approximation, successive transformations have been performed
which allow $\hat{J}(\bar{q})$ to be expressed in a suitable form
for our purposes. First the normalized moment $\bar{N}^{(2)}(\bar{q})$
defined above (Section \ref{OVKER}) is introduced in the definition
of $\hat{u}(\bar{q})$ which leads to \begin{eqnarray*}
\hat{u}(\bar{q}) & = & \frac{1}{\sqrt{N^{(0)}(\bar{q})}}\frac{1}{2}\bigl[N^{(2)}(\bar{q})P\bigr]^{(2)}\frac{1}{\sqrt{N^{(0)}(\bar{q})}}\\
 & = & -\frac{1}{2}C^{(2)}(N^{(2)},N^{(0)})+\frac{1}{2}\bigl[\bar{N}^{(2)}(\bar{q})P\bigr]^{(2)}.\end{eqnarray*}

The derivation of this expression is given in Appendix C. The coefficient
$C^{(2)}$ is a function of the first and second derivatives of the
moment of zero order. If the latter were constant, then $C^{(2)}$
would vanish. After inserting this expression in the definition of
$\hat{J}(\bar{q})$ given in equation (\ref{defu}) we obtain \begin{eqnarray*}
\hat{J}(\bar{q}) & = & 1+\alpha_{(0)}(\bar{q})+\frac{1}{2}\bigl[\bar{N}^{(2)}(\bar{q})P\bigr]^{(2)}\end{eqnarray*}
 with \begin{eqnarray*}
\alpha_{(0)}(\bar{q})=-\frac{1}{2}C^{(2)}(N^{(2)},N^{(0)}).\end{eqnarray*}

\begin{figure}[!h]

\begin{centering}
\includegraphics[width=8cm]{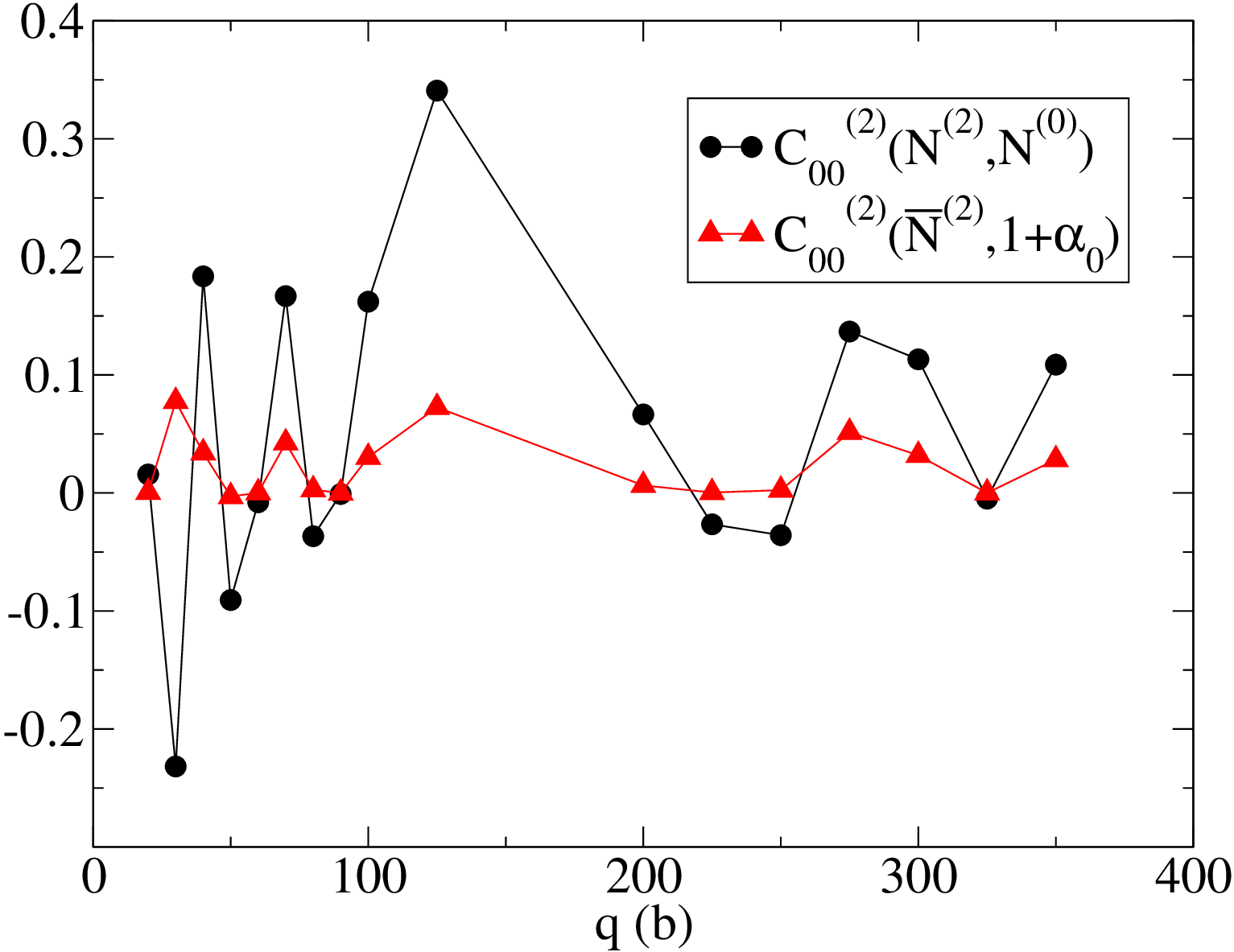} \caption{\label{alpha_0} Comparison between $C^{(2)}(N^{(2)},N^{(0)})$ and
first iterated $C^{(2)}(\bar{N}^{(2)},1+\alpha)$ values as a function
of the deformation.}

\par\end{centering}

\centering{}%
\end{figure}

According to Fig. \ref{alpha_0}, $\alpha_{(0)}(\bar{q})$ is not
negligible since it can reach $10\%$ at several deformations. It
is possible however to rewrite $\hat{J}(\bar{q})$ in the following
form \begin{eqnarray*}
\hat{J}(\bar{q})=\sqrt{A_{(0)}(\bar{q})}[1+\alpha_{(1)}(\bar{q})+\frac{1}{2}\bigl[\bar{N}_{R}^{(2)}(\bar{q})P\bigr]^{(2)}]\sqrt{A_{(0)}(\bar{q})}\end{eqnarray*}
 with \begin{eqnarray}
\begin{cases}
\alpha_{(1)}(\bar{q})=-\frac{1}{2}C^{(2)}(N^{(2)},A_{(0)})\\
\sqrt{A_{(0)}(\bar{q})}=\sqrt{1+\alpha_{(0)}(\bar{q})}\\
\bar{N}_{R}^{(2)}(\bar{q})=\frac{1}{\sqrt{A_{(0)}(\bar{q})}}\bar{N}^{(2)}\frac{1}{\sqrt{A_{(0)}(\bar{q})}}\end{cases}.\end{eqnarray}
 Coming back to Fig. \ref{alpha_0} we observe that $\alpha_{(1)}(\bar{q})$
is at most $4\%$ (at $q=30\,\textrm{b}$) in the whole range of deformation.
At this stage it seems reasonable to neglect this term but it is clear
that, if a better approximation is needed, one can iterate this procedure
starting with $\alpha_{(1)}$ which gives the quantity $A_{(1)}$
and so on. If one refers to Eqs. (\ref{calcJ2}) and (\ref{eq:EqSch2-ord2}),
one concludes in the light of this study that the entire formalism
remains unchanged with a $\hat{J}(\bar{q})$ still given by \begin{equation}
\hat{J}(\bar{q})=1+\hat{u}(\bar{q})=1+\frac{1}{2}\bigl[\bar{N}^{(2)}(\bar{q})P\bigr]^{(2)},\label{defureno}\end{equation}
 provided we substitute in all expressions a renormalized zero moment
$N_{R}^{(0)}(\bar{q})=\prod_{i=0}^{p-1}A_{(i)}N^{(0)}(\bar{q})$
instead of the usual moment $N^{(0)}(\bar{q})$. An index $i$ is
used to take into account the number of times the procedure described
above has been iterated. From now on the notation is kept unchanged,
with the understanding that in fact a renormalized zero moment is
used.

\begin{figure}[!h]

\begin{centering}
\includegraphics[width=8cm]{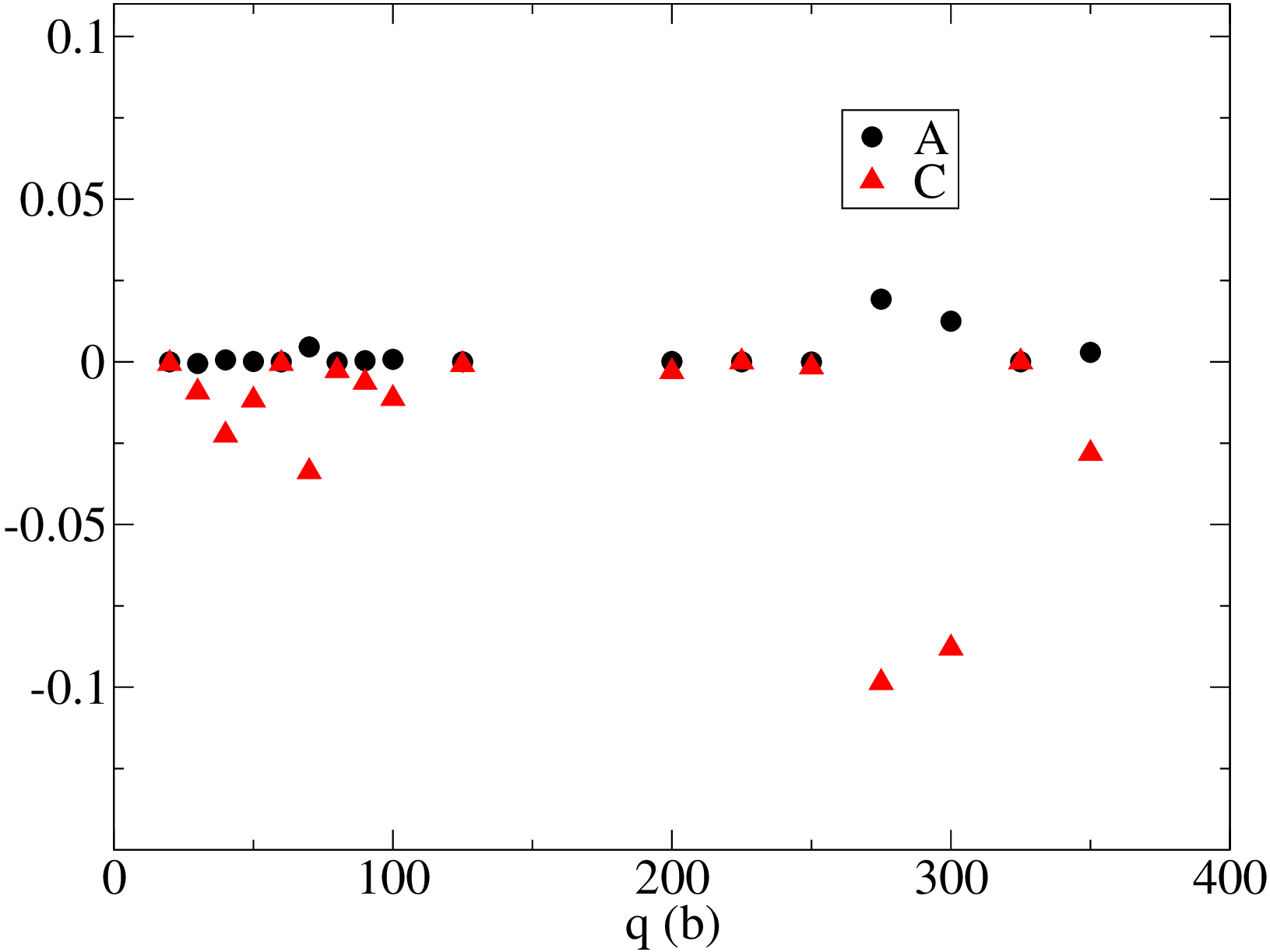} \caption{ \label{AC} $A$ and $C$ coefficients in Eq. (\ref{eq:remainder})
as function of the deformation.}

\par\end{centering}

\centering{}%
\end{figure}

Thus we are led to determine $\hat{J}_{\pm1/2}(\bar{q})$ with $\hat{J}(\bar{q})$
given in Eq. (\ref{defu}). In fact, we are mainly interested in the
determination of $\hat{J}_{-1/2}(\bar{q})$, since it is this operator
we need to express the contribution of the Hamiltonian kernel in Eq.
(\ref{eq:EqSch2-ord2}). In the one-dimensional case it must be solution
of the equation: \begin{equation}
\hat{J}_{-1/2}(\bar{q})[1+\frac{1}{2}\bigl[\bar{N}^{(2)}(\bar{q})P\bigr]^{(2)}]\hat{J}_{-1/2}(\bar{q})=1\label{eqJ-1/2}\end{equation}
 As suggested in reference \cite{RingSchuck80}, we attempt to solve
Eq. (\ref{eqJ-1/2}) with an ansatz of the form \begin{equation}
\hat{J}_{-1/2}(\bar{q})=A_{-1/2}(\bar{q})+\bigl[B_{-1/2}(\bar{q})P\bigr]^{(2)}+\bigl[C_{-1/2}(\bar{q})P\bigr]^{(4)}\end{equation}
 After inserting this ansatz in Eq. (\ref{eqJ-1/2}) we are led to
a complicated coupled set of non linear equations between $A_{-1/2}(\bar{q}),B_{-1/2}(\bar{q}),C_{-1/2}(\bar{q})$
and their derivatives up to high orders. Solving these equations,
including the second derivative of $N^{(2)}(\bar{q})$ is not a trivial
problem and it would be a considerable amount of work. This is an
extensive study in its own right that goes far beyond the scope of
the present paper. In the following we take advantage of the fact
that the size of the problem is reduced considerably if we neglect
these second derivatives and consequently we limit ourselves to finding
solutions including only the first derivative of $N^{(2)}(\bar{q})$.
Even in this simple case it is important to find approximated solutions
as a starting point in order to overcome some difficulties in solving
Eq. (\ref{eqJ-1/2}). A natural choice for these approximated solutions
is to approach $\hat{J}_{\pm1/2}(\bar{q})$ by the first terms of
their series expansion. For instance we will define a $\hat{J}_{A,1/2}(\bar{q})$
as: \[
\hat{J}_{A,1/2}(\bar{q})\equiv1+\frac{1}{2}\hat{u}(\bar{q})-\frac{1}{8}\hat{u}^{2}(\bar{q})\approx(1+\hat{u}(\bar{q}))^{1/2}\]
 where the operator $\hat{u}(\bar{q})$ is defined by Eq. (\ref{defureno}).
By inserting the definition of $u(\bar{q})$ in $\hat{J}_{A,1/2}(\bar{q})$
we readily obtain this operator in the form: \[
\hat{J}_{A,1/2}(\bar{q})=A_{1/2}(\bar{q})+\bigl[B_{1/2}(\bar{q})P\bigr]^{(2)}+\bigl[C_{1/2}(\bar{q})P\bigr]^{(4)}\]
 For the sake of brevity, the derivation of this expression and the
coefficients are given in Appendix \ref{C}2. At this stage we have
thought it interesting to study qualitatively the influence of the
derivatives of $\bar{N}^{(2)}(\bar{q})$ in this formalism. For this
we check the accuracy the approximation used by calculating $\hat{J}_{A,1/2}(\bar{q})\hat{J}_{A,1/2}(\bar{q})$
which we write in the form:

\[
\hat{J}_{A,1/2}(\bar{q})\hat{J}_{A,1/2}(\bar{q})=1+\hat{u}(\bar{q})+R(\bar{q}).\]
 The remainder $R(\bar{q})$, whose calculation is given in details
in the Appendix \ref{C}3 is in fact an operator \begin{equation}
R(\bar{q})\simeq A(\bar{N}^{(2)''}(\bar{q}))+\frac{1}{2}\bigl[C(\bar{N}^{(2)},\bar{N}^{(2)'},\bar{N}^{(2)''})\bar{N}^{(2)}(\bar{q})P\bigr]^{(2)}\label{eq:remainder}\end{equation}
 The symbol $\simeq$ is used to indicate that symmetric operators
$[.P]^{(n)},n>2$ are neglected in the expression of $R(\bar{q})$.
This operator, which should be zero if the moments were constant,
provides some quantitative information on the accuracy of our approximation
as function of the first and second derivatives of $\bar{N}^{(2)}(\bar{q})$.
In Fig. \ref{AC} we give a plot of the coefficients $A$ and $C$
over a wide range of deformations between $30\,\textrm{b}$ and $350\,\textrm{b}$.
It is seen that $\underset{\bar{q}}{\sup}|A|=1.9.10^{-2}$ and $\underset{\bar{q}}{\sup}|C|=9.8.10^{-2}$
over the whole range of deformations considered here, which suggests
that this ansatz is a reasonable zeroth-order approximation, despite
the large variations in the moments $\bar{N}^{(2)}$. On the other
hand, the expressions for the coefficients $A$ and $C$ given in
Appendix C tell us that $A$ and $C$ vanish independent if one can
neglect the second derivative of $\bar{N}^{(2)}(\bar{q})$. In that
case, $\hat{J}_{A,1/2}(\bar{q})$ becomes an \textquotedbl{}exact\textquotedbl{}
solution consistent with the truncation we made in the expansion of
the kernel overlap. Notice also that such an approximation takes into
account all the derivatives of $N^{(0)}(\bar{q})$ up to second order
and still includes the first derivative of $\bar{N}^{(2)}(\bar{q})$.

The same discussion applies to the operator $\hat{J}_{A,-1/2}(\bar{q})$
which is written in the form \[
\hat{J}_{A,-1/2}(\bar{q})=A_{-1/2}(\bar{q})+\bigl[B_{-1/2}(\bar{q})P\bigr]^{(2)}+\bigl[C_{-1/2}(\bar{q})P\bigr]^{(4)},\]
 the coefficients $A_{-1/2}(\bar{q}),B_{-1/2}(\bar{q}),C_{-1/2}(\bar{q})$
being given in Appendix \ref{C}2. In particular, if we neglect $\bar{N}^{(2)''}(\bar{q})$
this operator satisfies $\hat{J}_{A,-1/2}(\bar{q})\hat{J}_{A,+1/2}(\bar{q})\simeq1$
and more importantly it is also solution of Eq. (\ref{eqJ-1/2}) up
to second order. Inspection of the equations reveals that the coefficient
$C_{-1/2}(\bar{q})$ is not completely defined at this order. A complete
determination of this coefficient is achieved by canceling the contribution
of fourth order in Eq. (\ref{eqJ-1/2}). Its value is then given by
\begin{eqnarray*}
C_{-1/2}(\bar{q})=\frac{3}{32}(\bar{N}^{(2)}(\bar{q}))^{2}[1-\frac{15}{8}\xi(\bar{q})+\frac{105}{64}\xi^{2}(\bar{q})]\end{eqnarray*}
 with $\xi(\bar{q})=(\bar{N}^{(2)'}(\bar{q}))^{2}/\bar{N}^{(2)}(\bar{q})$.\\
 To obtain this expression, the third and fourth derivatives of
$C_{-1/2}(\bar{q})$ are set to zero. It turns out that these conditions
are satisfied by the solution given above since we set to zero the
derivatives $\left(\bar{N}^{(2)}(\bar{q})\right)^{(p)},p\ge2$. Thus
we conclude that it is a solution in the whole range of deformations
consistent with our assumptions on the derivatives of $\bar{N}^{(2)}(\bar{q})$.
To summarize, the operator \begin{eqnarray*}
\hat{J}_{-1/2}(\bar{q}) & = & 1-\frac{1}{4}\bigl[\bar{N}^{(2)}(\bar{q})(1-\frac{3}{8}(\bar{N}^{(2)'}(\bar{q}))^{2})P\bigr]^{(2)}\\
 & + & \frac{3}{32}\bigl[(\bar{N}^{(2)}(\bar{q}))^{2}(1-\frac{15}{8}\xi(\bar{q})+\frac{105}{64}\xi^{2}(\bar{q}))P\bigr]^{(4)}\end{eqnarray*}
 is solution of Eq. (\ref{eqJ-1/2}) up to forth order in the symmetric
operator $[.P]^{(n)}$. This expansion takes into account not only
the derivatives of $\bar{N}^{(0)}(\bar{q})$ up to second order but
also the first derivative of $\bar{N}^{(2)}(\bar{q})$. Even though
it is not complete, since we do not include the second derivative
of $\bar{N}^{(2)}(\bar{q})$, this operator represents a significant
improvement over what was generally used in previous works. We now
proceed to the $N$-dimensional case by following step by step the
procedure described in this section. Consequently, we will also neglect
the contribution of the second derivatives of $\bar{N}^{(2)}(\bar{q})$
in the expression of the inverse overlap kernel. One finds a justification
for this approximation in the fact that the general structure of the
Schrödinger equation we derive does not depend on the precise determination
of the coefficients entering in the definition of our ansatz for the
inverse.

\subsection{Determination of the inverse in the $N$-dimensional case}

A similar analysis to the one-dimensional case can be performed here
by considering instead the operator \begin{eqnarray*}
\hat{u}(\bar{q})=\frac{1}{\sqrt{N^{(0)}(\bar{q})}}(\bigl[N^{(1)}(\bar{q})P\bigr]^{(1)}+\frac{1}{2}\bigl[N^{(2)}(\bar{q})P\bigr]^{(2)})\frac{1}{\sqrt{N^{(0)}(\bar{q})}}.\end{eqnarray*}
 First it is written in terms of normalized moments according to the
expression \[
\hat{u}(\bar{q})=\bigl[W_{N}(\bar{q})P\bigr]^{(1)}+\frac{1}{2}\bigl[\bar N^{(2)}(\bar{q})P\bigr]^{(2)}+\alpha_{(0)}(\bar{q}),\]
 where we have introduced the quantities: \begin{eqnarray*}
W_{N}(\bar{q}) & = & \bar{N}^{(1)}(\bar{q})+iC^{(1)}(N^{(2)},N^{(0)})\\
\alpha_{(0)}(\bar{q}) & = & \frac{i}{2}C^{(1)}(N^{(1)},N^{(0)})-\frac{1}{2}C^{(2)}(N^{(2)},N^{(0)}).\end{eqnarray*}
 The expression of $C^{(1)}(N^{(1)},N^{(0)})$ is given in Appendix
\ref{C}1. According to the definition of $\hat{J}(\bar{q})$ (Eq.
(\ref{defu})) this operator becomes \[
\hat{J}(\bar{q})=I+\alpha_{(0)}(\bar{q})+\bigl[W_{N}(\bar{q})P\bigr]^{(1)}+\frac{1}{2}\bigl[\bar{N}^{(2)}(\bar{q})P\bigr]^{(2)}.\]
 The difference with the one-dimensional case is the occurrence of
an additional first-order symmetric operator and the fact that all
operators are matrices. At this stage we proceed as we did in the
one-dimensional case by introducing renormalized zero-order moments
and using an iterative procedure. Let us mention here that our numerical
result tell us that the matrix elements of $\alpha_{(0)}(\bar{q})$
never exceed $2\times10^{-1}$. Consequently a series expansion with
few terms gives the operator $(I+\alpha_{(0)}(\bar{q}))^{-1/2}$ with
excellent accuracy. We recall that we keep the same notation $N^{(0)}(\bar{q})$
to denote this renormalized moment. With these renormalizations we
are finally led to solve Eq. (\ref{eqJ-1/2}) which can be written
as \begin{eqnarray}
\hat{J}_{-1/2}(\bar{q})[I+\hat{u}(\bar{q})]\hat{J}_{-1/2}(\bar{q})=I\nonumber \\
\hat{u}(\bar{q})=\bigl[W_{N}(\bar{q})P\bigr]^{(1)}+\frac{1}{2}\bigl[\bar{N}^{(2)}(\bar{q})P\bigr]^{(2)}.\label{eqJ-1/2mat}\end{eqnarray}
 One could attempt to solve this equation with an appropriate ansatz
which, in the N-dimensional case, should also include odd-order symmetric
operators with an ansatz of the form \begin{eqnarray}
\hat{J}_{-1/2}(\bar{q}) & = & A_{-1/2}(\bar{q})+\bigl[D_{-1/2}(\bar{q})P\bigr]^{(1)}+\bigl[B_{-1/2}(\bar{q})P\bigr]^{(2)}\nonumber \\
 & + & \bigl[E_{-1/2}(\bar{q})P\bigr]^{(3)}+\bigl[C_{-1/2}(\bar{q})P\bigr]^{(4)}.\label{defJmat}\end{eqnarray}
 This expression should be inserted in Eq. (\ref{eqJ-1/2mat}) but
for reasons given in the one-dimensional case, it is advantageous
to start with an approximation to $\hat{J}_{A,-1/2}(\bar{q})$ chosen
as the series expansion of $(I+\hat{u}(\bar{q}))^{-1/2}$ up to second
order in $\hat{u}(\bar{q})$, i.e. we set \[
\hat{J}_{A,-1/2}(\bar{q})=I-\frac{1}{2}\hat{u}(\bar{q})+\frac{3}{8}\hat{u}^{2}(\bar{q}),\]
 with the operator $\hat{u}(\bar{q})$ defined this time by equation
(\ref{eqJ-1/2mat}). It is straightforward to express $\hat{J}_{-1/2}(\bar{q})$
in a form similar to Eq. (\ref{defJmat}) and to rewrite Eq. (\ref{defJmat})
as: \begin{eqnarray*}
\hat{J}_{-1/2}(\bar{q}) & = & I-\frac{1}{2}\bigl[W_{N}(\bar{q})P\bigr]^{(1)}+-\frac{1}{2}\bigl[(\frac{\bar{N}^{(2)}(\bar{q})}{2}\\
 & - & \frac{3}{16}(\bar{N}^{(2)'}(\bar{q}))^{2}-\frac{3}{4}(W_{N}(\bar{q}))^{2}\\
 & - & \frac{3i}{16}(W_{N}(\bar{q})\bar{N}^{(2)'}(\bar{q}))_{-})P\bigr]^{(2)}\\
 & + & \bigl[E_{-1/2}(\bar{q})P\bigr]^{(3)}+\bigl[C_{-1/2}(\bar{q})P\bigr]^{(4)}\end{eqnarray*}
 with the notation \[
(X,Y)_{\pm}=XY\pm YX.\]
 The first derivative of $W_{N}(\bar{q})$ has been neglected since
it is generally very small. The operator defined by the first three
lines of this expression is solution to Eq. (\ref{eqJ-1/2mat}) up
to second order. The coefficients $E_{-1/2}(\bar{q})$ and $C_{-1/2}(\bar{q})$
can then be determined by canceling the third and fourth order contribution
in Eq. (\ref{eqJ-1/2mat}). For practical reasons we neglect their
contributions and derive in the next section a Schrödinger equation
with the inverse defined by \begin{eqnarray}
\hat{J}_{-1/2}(\bar{q}) & = & I-\frac{1}{2}\bigl[W_{N}(\bar{q})P\bigr]^{(1)}+-\frac{1}{2}\bigl[(\frac{\bar{N}^{(2)}(\bar{q})}{2}\nonumber \\
 & - & \frac{3}{16}(\bar{N}^{(2)'}(\bar{q}))^{2}-\frac{3}{4}(W_{N}(\bar{q}))^{2}\nonumber \\
 & - & \frac{3i}{16}(W_{N}(\bar{q})\bar{N}^{(2)'}(\bar{q}))_{-})P\bigr]^{(2)}.\label{Jred}\end{eqnarray}
 Otherwise the expression would have contained an exceedingly large
number of terms. Again we justify this approximation by the fact that
it does not affect the general structure of the Schrödinger equation
that we propose.

\section{Derivation of a Schrödinger-like equation}

\label{SE}

According to Eq. (\ref{eq:EqSch2-ord2}) and taking into account the
fact that the inverse operator $\hat{J}_{-1/2}(\bar{q})$ given by
Eq. (\ref{Jred}) is Hermitian, the contribution of the Hamiltonian
kernel to the Hill-Wheeler equation reduces to the form

\begin{eqnarray*}
\hat{J}_{-1/2}(\bar{q})\frac{1}{\sqrt{N^{(0)}(\bar{q})}}(H^{(0)}(\bar{q})+\bigl[H^{(1)}(\bar{q})P\bigr]^{(1)}\\
+\frac{1}{2}\bigl[H^{(2)}(\bar{q})P\bigr]^{(2)})\frac{1}{\sqrt{N^{(0)}(\bar{q})}}\hat{J}_{-1/2}(\bar{q}).\end{eqnarray*}
 Now, we proceed as we did in the case of the overlap kernel by introducing
this time normalized moments of the Hamiltonian defined by \[
\bar{H}^{(n)}(\bar{q})=\frac{1}{\sqrt{N^{(0)}(\bar{q})}}H^{(n)}(\bar{q})\frac{1}{\sqrt{N^{(0)}(\bar{q})}},\]
 and rewrite the equation given above as function of these moments
\begin{eqnarray}
 &  & \hat{J}_{-1/2}(\bar{q})(\bar{H}^{(0)}(\bar{q})+\frac{i}{2}C^{(1)}(\bar{H}^{(1)},\bar{N}^{(0)})\nonumber \\
 & - & \frac{1}{2}C^{(2)}(\bar{H}^{(1)},\bar{N}^{(0)})+\bigl[(\bar{H}^{(1)}(\bar{q})+\frac{i}{2}C^{(1)}(\bar{H}^{(2)},\bar{N}^{(0)}))P\bigr]^{(1)}\nonumber \\
 & + & \frac{1}{2}\bigl[\bar{H}^{(2)}(\bar{q})P\bigr]^{(2)})\hat{J}_{-1/2}(\bar{q}).\label{contH}\end{eqnarray}
 The next step is to extract a Schrödinger-like equation following
the procedure described in \cite{RingSchuck80}. To this end we insert
the definition from Eq. (\ref{Jred}) in Eq. (\ref{contH}) and make
an expansion in terms of symmetric ordered product that we truncate
at order two. We illustrate in the following how we achieve this goal
by rewriting Eq. (\ref{contH}) in condensed notation \[
\hat{J}_{-1/2}(\bar{q})\mathcal{H}(\bar{q})\hat{J}_{-1/2}(\bar{q}),\]
 with \begin{eqnarray}
\begin{cases}
\mathcal{H}(\bar{q})=\sum_{i=0}^{2}\bigl[h_{(i)}(\bar{q})P\bigr]^{(i)}\\
\hat{J}_{-1/2}(\bar{q})=\sum_{i=0}^{2}\bigl[j_{(i)}(\bar{q})P\bigr]^{(i)}\end{cases}\end{eqnarray}
 where the quantities $j_{(i)}(\bar{q}),h_{(i)}(\bar{q})$ are obtained
by identification of these operators with those defined in Eq. (\ref{Jred})
and Eq. (\ref{contH}) respectively. With these notations we can rewrite
Eq. (\ref{contH}) in the convenient form \[
\sum_{p,q,r=0}^{2}\bigl[j_{(p)}(\bar{q})P\bigr]^{(p)}\bigl[h_{(q)}(\bar{q})P\bigr]^{(q)}\bigl[j_{(r)}(\bar{q})P\bigr]^{(r)},\]
 which has the advantage of showing the typical terms occurring in
this formalism. The expansion we are looking for is then derived very
simply by means of general formulas that are given in Appendix \ref{A}
(Eqs. (\ref{SOPO2}) and (\ref{eq:SOPO3})). This derivation is straightforward,
but requires nevertheless quite cumbersome manipulations that we do
not describe here. Thus we limit ourselves to giving here the final
expression of a Schrödinger equation which takes into account the
coupling between collective and intrinsic excitations. This equation
is written as \begin{eqnarray}
 &  & \left(H(\bar{q})-E\right)g(\bar{q})=0,\nonumber \\
 & H & (\bar{q})=S(\bar{q})+\bigl[T(\bar{q})P\bigr]^{(1)}+\bigl[U(\bar{q})P\bigr]^{(2)}\label{HSTU}\end{eqnarray}
 where the quantities $S$, $T$, and $U$ are matrices whose expression
is \begin{eqnarray}
S & = & h_{(0)}+\frac{1}{8}(h_{(0)}^{(2)}j_{(2)})_{+}+\frac{1}{16}j_{(2)}^{(1)}h_{(0)}^{(2)}j_{(2)}^{(1)}-\frac{1}{4}(h_{(0)}^{(1)}j_{(1)})_{-}\nonumber \\
 & - & \frac{1}{16}j_{(1)}h_{(0)}^{(2)}j_{(1)}+\frac{1}{32}(j_{(1)}h_{(0)}^{(2)}j_{(2)}^{(1)})_{A}\nonumber \\
T & = & h_{(1)}+\frac{1}{2}(h_{(0)}j_{(1)})_{+}-\frac{1}{2}(h_{(0)}^{(1)}j_{(2)})_{-}-\frac{1}{16}(j_{(1)}h_{(0)}^{(2)}j_{(2)})_{S}\nonumber \\
 & - & \frac{1}{8}(j_{(1)}h_{(0)}^{(1)}j_{(1)}^{(1)})_{S}-\frac{1}{16}(j_{(2)}h_{(0)}^{(2)}j_{(2)}^{(1)})_{A}\nonumber \\
U & = & h_{(2)}+\frac{1}{2}(h_{(0)}j_{(2)})_{+}-\frac{1}{2}(h_{(2)}^{(1)}j_{(2)}^{(1)})_{+}-\frac{1}{4}j_{(2)}^{(1)}h_{(0)}j_{(2)}^{(1)}\nonumber \\
 & - & \frac{1}{8}j_{(2)}h_{(0)}^{(2)}j_{(2)}-\frac{1}{8}(j_{(2)}^{(1)}h_{(0)}^{(1)}j_{(2)})_{S}+\frac{1}{2}(h_{(1)}j_{(1)})_{+}\nonumber \\
 & + & \frac{1}{4}(h_{(1)}j_{(2)}^{(1)})_{-}-\frac{1}{4}(h_{(2)}^{(1)}j_{(1)})_{-}+\frac{1}{4}j_{(1)}h_{(0)}j_{(1)}\nonumber \\
 & + & \frac{1}{8}(j_{(2)}h_{(0)}^{(1)}j_{(1)})_{A}-\frac{1}{8}(j_{(2)}^{(1)}h_{(0)}j_{(1)})_{A}\label{STU}\end{eqnarray}
 For the sake of convenience, the dependence in $\bar{q}$ has been
removed in all the quantities in this formula where we use the compact
notation \begin{eqnarray*}
(XYZ)_{S} & = & XYZ+ZYX\\
(XYZ)_{A} & = & XYZ-ZYX.\end{eqnarray*}

Furthermore, the expression for the matrix elements of the moments
of $H(\bar{q})$ suggest that we make the same assumptions as for
the derivatives of the overlap moments. Thus in the derivation of
the formulas proposed here it has been assumed that: \begin{eqnarray}
\begin{cases}
h_{(0)}^{(p)}(\bar{q})=0\hspace{1cm}\text{for}\hspace{0.5cm}p\ge3\\
h_{(1)}^{(p)}(\bar{q})=0\hspace{1cm}\text{for}\hspace{0.5cm}p\ge1\\
h_{(2)}^{(p)}(\bar{q})=0\hspace{1cm}\text{for}\hspace{0.5cm}p\ge2\end{cases}\end{eqnarray}

These assumptions serve to reduce significantly the number of terms
in the expressions of the different contributions $S(\bar{q}),T(\bar{q}),U(\bar{q})$.

We discuss now some aspects related to the properties and the general
structure of the Schrödinger equation defined by Eq. (\ref{HSTU})
and Eq. (\ref{STU}) which contains all the necessary ingredients
to build it. First, we note that this Hamiltonian is Hermitian and
time-reversal invariant. This is easily demonstrated using the properties
of the matrix elements of the moments listed in Section \ref{anamom}.
As a consequence, if we treat collective and intrinsic excitations
on the same footing, there is no dissipation but only a reversible
exchange of energy between these two kinds of excitation. Notice in
particular that a linear term as it occurs in Eq. (\ref{HSTU}) does
not induce a dissipative force as introduced in studies of fission
using a macroscopic-microscopic approach \cite{Carjan86}. However,
It may be that for practical reasons we need to treat explicitly a
reduced set of intrinsic excitations and take into account the remainder
implicitly because they are too numerous and their spectrum too dense.
In that case we may be led to introduce a dissipation mechanism. We
do not consider for the moment applications at high excitation energies
where such situations would occur \cite{Dietrich10}. Concerning the
general structure of the Schrödinger Eq. (\ref{HSTU}), we analyze
separately the three different terms which are present in its definition.
In Eq. (\ref{STU}) we have arranged the terms in such a way that
even moments come first and odd ones at the end in $S(\bar{q})$ and
$U(\bar{q})$. We recall that odd moments occur only in the multi-dimensional
case. Accordingly in evaluating $S(\bar{q})$ and $U(\bar{q})$ in
the one-dimensional case we must consider only the first three terms
in $S(\bar{q})$ and the first six in $U(\bar{q})$. Their expressions
are readily obtained with the formulas given previously but they are
very long and for the sake of brevity we do not give them here. Notice
that if we neglect the derivatives of the overlap moments we recover,
as we should, the simple Hamiltonian given in reference \cite{RingSchuck80}.
In the multi-dimensional case all the terms in Eq. (\ref{STU}) must
be retained. They involve products of matrices of the moments of the
overlap and Hamiltonian kernels, and their derivatives. Each of them
can be interpreted as effective vertices inducing transitions between
excitations. At this point it is important to note that these vertices,
by the definition of the moments, are the result of an average over
a wide range of deformations around $\bar{q}$ and consequently the
coupling between collective and intrinsic degrees of freedom is very
non-local in the present work. This contrasts with the other approach
\cite{Moya77} which couples the nuclear collective dynamics to internal
excitations defined at every deformation along the adiabatic fission
path. To conclude this analysis it is appropriate to recast our Hamiltonian
Eq. (\ref{HSTU}) in a familiar form. With more conventional notations,
first we express its diagonal matrix elements which leads to the expression
\[
H_{ii}(\bar{q})=\frac{1}{2}\bigl[(\frac{1}{M(\bar{q})})_{ii}P\bigr]^{(2)}+V_{ii}(\bar{q}),\]

We have taken into account the fact that the matrix $T$ is zero along
the diagonal and identified a collective potential as: $V(\bar{q})=S(\bar{q})$.
Also we have defined the mass as: $M^{-1}(\bar{q})=2U(\bar{q})$.
We recognize standard Hamiltonians which describe separately the collective
dynamics on each potential energy surfaces built with the various
2-qp excitations considered. In the same manner we introduce the coupling
between 2-qp surfaces by considering the non-diagonal matrix elements
of Eq. (\ref{HSTU}). They are written in the form \[
H_{ij}(\bar{q})=\frac{1}{2}\bigl[\left(\frac{1}{M(\bar{q})}\right)_{ij}P\bigr]^{(2)}+\bigl[T_{ij}(\bar{q})P\bigr]^{(1)}+V_{ij}(\bar{q}),\]
 with the definition of a non-diagonal mass as: $\left(\frac{1}{M(\bar{q})}\right)_{ij}=2U_{ij}(\bar{q})$.

This form of the interaction term is interesting because it separates
the coupling between collective and intrinsic degrees of freedoms
into terms of different nature. The first two, which depend on the
collective momentum, can be interpreted as a dynamical coupling while
the last contribution is a potential coupling. In appearance the Hamiltonian
derived here has a general structure very similar to the classical
Hamiltonian given by the authors in Ref. \cite{Moya77} . However
their derivation and ours rely on very different approaches. In particular
the one presented here is based on a quantum description from the
start, while the classical Hamiltonian in Ref. \cite{Moya77} needs
to be quantized after the fact.

For completeness, we could have included in our discussion the explicit
calculation of the moments associated with neutron- and proton-number
operators, in order to use the method \cite{Bonche90} which incorporates
the conservation of the average particle number in the GCM formalism.
Similarly, we could have included the moments of the operator associated
with the subtraction of the center-of-mass energy of the nucleus.
These contributions are straightforward to calculate, but are omitted
here in order to avoid further complicating the expression for the
Hamiltonian above.

\section{Conclusion}

\label{concl}

In this paper we have presented a theoretical framework, the Schrödinger
Collective Intrinsic Model (SCIM), which allows in a microscopic way
the simultaneous coupling of single-particle and collective degrees
of freedom. Such an approach is based on a generalized GCM, where
the general GCM ansatz of the nuclear wave function is extended by
a few excited configurations. In fact, one considers as generating
wave functions not only HFB ground-state configurations with different
values for the collective generator coordinate but also 2-qp excited
states. Such an approach has the advantage of describing in a completely
quantum-mechanical fashion, and without phenomenological parameters,
the coupling of qp degrees of freedom to the collective motion of
the nucleons. Our derivation of the SCIM proceeds in two steps. The
first step is based on the generalization of the symmetric moment
expansion of the equations derived in the framework of the GCM including
the coupling between collective and intrinsic variables. The moments
occurring in the equations are matrices whose dimensions depends on
the number of excitations introduced in the description. Contrary
to the usual case -- without qp excitations -- all moments, odd and
even, must be included in the summation. Such an exact expansion in
terms of local operators of the generalized Hill-Wheeler equation
is then transformed into a local Schrödinger equation by inverting
the expansion of the overlap kernel. Then a second-order differential
Schrödinger equation is derived by limiting the summation to second
order terms under the assumption that the series expansion of the
overlap and Hamiltonian kernels converge rapidly. The derivation of
the inverse of the overlap has been seriously complicated by the fact
that the moments and their derivatives vary rapidly as function of
the deformation, as observed in the numerical study of the overlap
kernel in a wide range of deformations in $^{236}$U. A quantitative
information on the accuracy of our approximation as function of the
first and second derivatives of second-order moment of the overlap
has been given in the present paper for the scalar case. The Schrödinger
equation is finally written in a convenient form that has the advantage
of exhibiting typical terms occurring in the formalism such as, for
instance, the potential coupling, and the dynamical coupling between
intrinsic and collective excitations. There are a number of avenues
that could be pursued now with this new formalism. Among them, the
study of the fission dynamics and the coupling between intrinsic and
collective excitations in the descent from saddle to scission represents
one of the most challenging problems in many-body theory. Also, a
particularly interesting use for this new approach is in those nuclei
which exhibit excitation spectra showing collective as well as non-collective
effects. This would surely improve the energies of the $0_{2}^{+}$
states, which is one of the problems of collective models today. In
all these studies, the microscopic nuclear wave function will be analyzed
to determine the influence of the different degrees of freedom on
nuclear properties. This would of course add to the computational
burden, but the ingredients to perform the calculation are ready for
the most part. The derivation and calculation of energy kernels will
be presented in detail in a separate publication. 

\begin{acknowledgments}
The authors would like to warmly thank J.F. Berger and N. Dubray for
enlightening discussions and for very useful advice for the computational
part. This work was performed in part under the auspices of the US
Department of Energy by the Lawrence Livermore National Laboratory
under Contract DE-AC52-07NA27344. Funding for this work was provided
in part by the United States Department of Energy Office of Science,
Nuclear Physics Program pursuant to Contract DE-AC52-07NA27344 Clause
B-9999, Clause H-9999 and the American Recovery and Reinvestment Act,
Pub. L. 111-5. 
\end{acknowledgments}
\appendix

\section{SYMMETRIC ORDERED PRODUCT OF OPERATORS}

\label{A}

The symmetric ordered product of operators (SOPO) is defined by \cite{RingSchuck80}
(p.421): \begin{eqnarray}
\bigl[A(q+s/2,q-s/2)P\bigr]^{(n)}=\nonumber \\
\frac{1}{2^{n}}\sum_{q=0}^{n}C_{n}^{q}P^{n-q}A(q+s/2,q-s/2)P^{q}\label{SOPO}\end{eqnarray}
 where $A$ is any operator and $P$ is the derivative operator $P=i\frac{\partial}{\partial\bar{q}}$.
For instance, the first three SOPO are: \begin{eqnarray*}
\bigl[AP\bigr]^{(0)} & = & A\\
\bigl[AP\bigr]^{(1)} & = & \frac{1}{2}(AP+PA)\\
\bigl[AP\bigr]^{(2)} & = & \frac{1}{4}(AP^{2}+2PAP+P^{2}A)\end{eqnarray*}
 where the action of $PA$ on a vector $g(q)$ is: \begin{equation}
PAg(q)=A^{(1)}g(q)+APg(q)\end{equation}
 with $A^{(n)}$ defined as $A^{(n)}=(P^{n}A)=(i)^{n}\frac{\partial^{n}A}{\partial q^{n}}$.
Let us note that for a Hermitian operator $A$, $A^{(n)}$ is Hermitian
or anti-Hermitian according to the parity of $n$: \[
\left(A^{(n)}\right)^{+}=(1)^{n}A^{(n)}\]
 For any operators $A$, $B$, and $C$, which depend on the collective
variable $q$, we have the following properties:\\
 - The symmetric ordered product is linear: \[
\bigl[AP\bigr]^{(n)}+\bigl[BP\bigr]^{(n)}=\bigl[(A+B)P\bigr]^{(n)}\]
 - The symmetric ordered product preserves the hermiticity: \begin{eqnarray*}
A^{+}=A & \Rightarrow & \left(\bigl[AP\bigr]^{(n)}\right)^{+}=\bigl[AP\bigr]^{(n)}\\
\begin{cases}
A^{+}=A\\
B^{+}=B\end{cases} & \Rightarrow & \left(\bigl[AP\bigr]^{(n)}\bigl[BP\bigr]^{(q)}\right)^{+}=\bigl[BP\bigr]^{(q)}\bigl[AP\bigr]^{(n)}\end{eqnarray*}
 - More generally, the product of two and three SOPO can be expanded
as a linear combination of symmetric ordered product. This property
is used throughout the paper in the derivation of various expressions.

We give these expansions in a form convenient for retrieving the contribution
of a SOPO of a given order \begin{eqnarray}
\bigl[AP\bigr]^{(n)}\bigl[BP\bigr]^{(q)} & = & \sum_{i=0}^{n+q}\frac{1}{2^{i}}\sum_{s=\max(0,i-q)}^{\min(n,i)}C_{n}^{s}C_{q}^{i-s}(-1)^{i-s}\nonumber \\
 & \bigl[ & A^{(i-s)}B^{(s)}P\bigr]^{(n+q-i)}\label{SOPO2}\end{eqnarray}
 \begin{eqnarray}
\bigl[AP\bigr]^{(n)}\bigl[BP\bigr]^{(q)}\bigl[CP\bigr]^{(r)}=\sum_{p=0}^{n+q+r}\frac{(-1)^{p}}{2^{p}}\sum_{i=0}^{\min(p,q+r)}\sum_{t=\max(0,p-q-r)}^{\min(p-i,n)}\nonumber \\
C_{n}^{t}C_{q+r-i}^{p-i-t}(-1)^{t}\bigl[A^{(p-i-t)}F(B,C,r,i,t)P\bigr]^{(n+q+r-p)}\label{eq:SOPO3}\end{eqnarray}
 with \[
F(B,C,r,i,t)=\sum_{s=\max(0,i-r)}^{\min(i,q)}(-1)^{s}C_{q}^{s}C_{r}^{i-s}\left(B^{(i-s)}C^{(s)}\right)^{(t)}.\]
 In particular, using (\ref{eq:SOPO3}) with $n=0$, $q=1$, $r=0$
we find: \begin{eqnarray}
A\bigl[BP\bigr]^{(1)}C & = & \bigl[ABCP\bigr]^{(1)}+\frac{1}{2}\left(ABC^{(1)}-A^{(1)}BC\right)\nonumber \\
\label{SOPO4}\\A\bigl[BP\bigr]^{(2)}C & = & \bigl[ABCP\bigr]^{(2)}+\bigl[(ABC^{(1)}-A^{(1)}BC)P\bigr]^{(1)}\nonumber \\
 & + & \frac{1}{4}\left(A^{(2)}BC-2A^{(1)}BC^{(1)}+ABC^{(2)}\right)\nonumber \end{eqnarray}

Formulas (\ref{SOPO4}) and (\ref{eq:SOPO3}) are the formulas used
to develop the formalism in terms of normalized moments. In fact after
setting \[
A=C=\frac{1}{\sqrt{N^{(0)}}}\]
 we obtain \begin{eqnarray}
\frac{1}{\sqrt{N^{(0)}}}\bigl[BP\bigr]^{(1)}\frac{1}{\sqrt{N^{(0)}}} & = & \bigl[\bar{B}P\bigr]^{(1)}+\frac{i}{2}C^{(1)}(B,N^{(0)})\nonumber \\
\frac{1}{\sqrt{N^{(0)}}}\bigl[BP\bigr]^{(2)}\frac{1}{\sqrt{N^{(0)}}} & = & \bigl[\bar{B}P\bigr]^{(2)}+iC^{(1)}(B,N^{(0)})\nonumber \\
 & - & C^{(2)}(B,N^{(0)})\label{eq:SOPO5}\end{eqnarray}
 with the definitions \begin{eqnarray*}
C^{(1)}(A,N^{(0)}) & = & \frac{1}{\sqrt{N^{(0)}}}A\left(\frac{1}{\sqrt{N^{(0)}}}\right)'-\left(\frac{1}{\sqrt{N^{(0)}}}\right)'A\frac{1}{\sqrt{N^{(0)}}}\\
C^{(2)}(A,N^{(0)}) & = & \frac{1}{4}\Biggl[\left(\frac{1}{\sqrt{N^{(0)}}}\right)''A\frac{1}{\sqrt{N^{(0)}}}\\
 & + & 2\left(\frac{1}{\sqrt{N^{(0)}}}\right)'A\left(\frac{1}{\sqrt{N^{(0)}}}\right)'\\
 & + & \frac{1}{\sqrt{N^{(0)}}}A\left(\frac{1}{\sqrt{N^{(0)}}}\right)''\Biggr]\end{eqnarray*}

\section{OVERLAP KERNEL}

\label{B}

The matrix elements of the overlap kernel are expressed as $N_{ij}(q,q')=\bra{\Phi_{i}(q)}\ket{\Phi_{j}(q')}$,
where $\ket{\Phi_{i}(q)}$ is defined by Eq. (\ref{defHFBi}).

The calculation of the overlap kernel between two HFB states at different
deformations is detailed in \cite{Haider89,Haider92}. We use the
formalism in \cite{Haider89,Haider92} because we are most familiar
with it, but of course the same formulas exist elsewhere in the literature
(e.g., \cite{RingSchuck80,Didong73,Bonche90} and references therein).
The generalization to 2-qp excitations is straightforward.

Let us introduce the quantities \textquotedbl{}$S$, $T$ and $Y$\textquotedbl{}
of \cite{Haider89,Haider92} \begin{eqnarray}
S_{ij} & = & \frac{\bra{\Phi(q)}|a_{j}^{+q'}a_{i}^{q'}\ket{\Phi(q')}}{\bra{\Phi(q)}\ket{\Phi(q')}},\nonumber \\
T_{ij} & = & \frac{\bra{\Phi(q)}|a_{i}^{+q'}a_{j}^{+q'}\ket{\Phi(q')}}{\bra{\Phi(q)}\ket{\Phi(q')}},\nonumber \\
Y_{ij} & = & \frac{\bra{\Phi(q)}|a_{i}^{q'}a_{j}^{q'}\ket{\Phi(q')}}{\bra{\Phi(q)}\ket{\Phi(q')}}\end{eqnarray}
 where the particle operators $\{a_{i}^{q},a_{i}^{+q}\}$ are related
to the qp operators $\{\eta_{i}^{q},\eta_{i}^{+q}\}$ through the
BCS transformation: \begin{equation}
\begin{cases}
\eta_{i}^{q}=u_{i}^{q}a_{i}^{q}-v_{i}^{q}a_{\bar{i}}^{+q}\\
\eta_{\bar{i}}^{+q}=u_{i}^{q}a_{\bar{i}}^{+q}+v_{i}^{q}a_{i}^{q}\end{cases}\end{equation}
 For $i,j,k,l>0$, and $\bar{i},\bar{j},\bar{k},\bar{l}$ their time
reversed states, we have: \begin{eqnarray}
\bra{\Phi(q)}|\eta_{i}^{+q'}\eta_{\bar{j}}^{+q'}\ket{\Phi(q')} & = & \bra{\Phi(q)}\ket{\Phi(q')}\bigl[u_{i}^{q'}u_{j}^{q'}T_{i\bar{j}}+u_{i}^{q'}v_{j}^{q'}S_{ji}\nonumber \\
 & - & v_{i}^{q'}u_{j}^{q'}(\delta_{ij}-S_{ij})+v_{i}^{q'}v_{j}^{q'}Y_{j\bar{i}}\bigr)],\nonumber \\
\bra{\Phi(q)}|\eta_{\bar{l}}^{q}\eta_{k}^{q}\ket{\Phi(q')} & = & \bra{\Phi(q)}\ket{\Phi(q')}\sum_{pp'}\tau_{lp}^{qq'}\tau_{kp'}^{qq'}\Bigl[u_{l}^{q}u_{k}^{q}Y_{\bar{p}p'}\nonumber \\
 & - & u_{l}^{q}v_{k}^{q}(\delta_{pp'}-S_{pp'})+v_{l}^{q}u_{k}^{q}S_{p'p}-v_{l}^{q}v_{k}^{q}T_{p\bar{p}'}\Bigr]\end{eqnarray}
 where the transformation $\tau_{ij}^{qq'}$ is defined through the
relation \begin{equation}
a_{j}^{+q'}=\sum_{i}\tau_{ij}^{qq'}a_{i}^{+q}\end{equation}

The different quantities in Eq. (B3) can be expressed as \cite{Haider89,Haider92}:
\begin{eqnarray}
\bra{\Phi(q)}\ket{\Phi(q')} & = & det(\tau^{qq'})det(Z)\nonumber \\
\delta_{ij}-S_{ij} & = & (\tau^{qq'-1}u^{q}Z^{'-1}u^{q'})_{ij}\nonumber \\
T_{i\bar{j}} & = & (u^{q'}Z^{-1}v^{q}\tau^{qq'})_{ji}\nonumber \\
Y_{i\bar{j}} & = & -(\tau^{qq'-1}u^{q}Z^{'-1}v^{q'})_{ij}\end{eqnarray}
 with $Z=u^{q}(\tau^{+})^{-1}u^{q'}+v^{q}\tau v^{q'}$ and $Z'=u^{q'}(\tau)^{-1}u^{q}+v^{q'}\tau^{+}v^{q}$.

\section{Calculation of $I+\hat{u}(q)$ and related quantities in terms of
normalized moments}

\label{C}

\subsection{Calculation of $I+\hat{u}(q)$}

According to Eq.(\ref{calcJ2}), the operator $I+\hat{u}(q)$ is \begin{eqnarray}
I+\hat{u}(q) & = & I+\frac{1}{\sqrt{N^{(0)}(q)}}(\bigl[N^{(1)}(q)P\bigr]^{(1)}\nonumber \\
 & + & \frac{1}{2}\bigl[N^{(2)}(q)P\bigr]^{(2)})\frac{1}{\sqrt{N^{(0)}(q)}}\end{eqnarray}
 By use of Eqs. (\ref{SOPO4}) and (\ref{eq:SOPO5}), $\hat{u}(q)$
can be expressed, after straightforward calculations, as \begin{equation}
\hat{u}(q)=C_{N}^{+}(q)+\bigl[W_{N}(q)P\bigr]^{(1)}+\frac{1}{2}\bigl[\bar{N}^{(2)}(q)P\bigr]^{(2)}\end{equation}
 with \begin{eqnarray*}
W_{N}(q) & = & \bar{N}^{(1)}(q)+iC^{(1)}(N^{(2)},N^{(0)})\\
C_{N}^{+}(q) & = & \frac{i}{2}C^{(1)}(N^{(1)},N^{(0)})-\frac{1}{2}C^{(2)}(N^{(2)},N^{(0)})\end{eqnarray*}
 and \begin{eqnarray*}
C^{(1)}(A,N^{(0)}) & = & \frac{1}{\sqrt{N^{(0)}(q)}}A\left(\frac{1}{\sqrt{N^{(0)}(q)}}\right)'\\
 & - & \left(\frac{1}{\sqrt{N^{(0)}(q)}}\right)'A\frac{1}{\sqrt{N^{(0)}(q)}}\\
C^{(2)}(A,N^{(0)}) & = & \frac{1}{4}\Biggl[\left(\frac{1}{\sqrt{N^{(0)}(q)}}\right)''A\frac{1}{\sqrt{N^{(0)}(q)}}\\
 & + & 2\left(\frac{1}{\sqrt{N^{(0)}(q)}}\right)'A\left(\frac{1}{\sqrt{N^{(0)}(q)}}\right)'\\
 & + & \frac{1}{\sqrt{N^{(0)}(q)}}A\left(\frac{1}{\sqrt{N^{(0)}(q)}}\right)''\Biggr]\end{eqnarray*}

It is worth noticing that in the scalar case, the coefficient $C^{(1)}(A,N^{(0)})$
and $\bar{N}^{(1)}(q)$ are zero so that the operator u reduces to
\[
\hat{u}(q)=-\frac{1}{2}C^{(2)}(N^{(2)},N^{(0)})+\frac{1}{2}\bigl[\bar{N}^{(2)}(q)P\bigr]^{(2)}\]

\subsection{Calculation of the operators $(I+\hat{u}(\bar{q}))^{1/2}$ and $(I+\hat{u}(\bar{q}))^{-1/2}$}

For sake of simplicity, we restrict our discussion to the standard
--one dimensional-- case. $\hat{J}_{1/2}(\bar{q})$ the square root
of the operator $1+\hat{u}(\bar{q})$ is approximated by the first
terms of the series expansion up to the order $\hat{u}^{2}(\bar{q})$:
\begin{eqnarray}
\hat{J}_{1/2}(\bar{q})=(I+\hat{u}(\bar{q}))^{1/2}=1+\frac{1}{2}\hat{u}(\bar{q})-\frac{1}{8}\hat{u}^{2}(\bar{q})\label{expJ1/2}\end{eqnarray}
 with \begin{eqnarray*}
\hat{u}(\bar{q})=\frac{1}{2}\bigl[\bar{N}^{(2)}(\bar{q})P\bigr]^{(2)}\end{eqnarray*}
 Using the property (A.1), $(\hat{u}(\bar{q}))^{2}$ is \begin{eqnarray*}
\hat{u}^{2}(\bar{q}) & = & \frac{1}{4}\Bigl[\frac{1}{16}\left(\bar{N}^{(2)''}(\bar{q})\right)^{2}+\frac{1}{2}\bigl[\Bigl(\bar{N}^{(2)''}(\bar{q})\bar{N}^{(2)}(\bar{q})\\
 & - & 2(\bar{N}^{(2)'}(\bar{q}))^{2}\Bigr)P\bigr]^{(2)}+\bigl[(\bar{N}^{(2)}(\bar{q}))^{2}P\bigr]^{(4)}\Bigr]\end{eqnarray*}
 and the operator $\hat{J}_{1/2}(\bar{q})$ can be expressed in terms
of SOPO as: \begin{equation}
\hat{J}_{1/2}(\bar{q})=A_{1/2}(\bar{q})+\bigl[B_{1/2}(\bar{q})P\bigr]^{(2)}+\bigl[C_{1/2}(\bar{q})P\bigr]^{(4)}\end{equation}
 with \begin{eqnarray}
\begin{cases}
A_{1/2}(\bar{q})=1-\frac{1}{2^{9}}\left(\bar{N}^{(2)''}(\bar{q})\right)^{2}\\
B_{1/2}(\bar{q})=\frac{1}{4}\left(1+\frac{1}{16}\bar{N}^{(2)''}(\bar{q})-\frac{1}{8}\frac{\left(\bar{N}^{(2)'}(\bar{q})\right)^{2}}{\bar{N}^{(2)}(\bar{q})}\right)\bar{N}^{(2)}(\bar{q})\\
C_{1/2}(\bar{q})=-\frac{1}{32}\left(\bar{N}^{(2)}(\bar{q})\right)^{2}\end{cases}\end{eqnarray}
 The same derivation applies for $\hat{J}_{-1/2}(\bar{q})$ the inverse
of the root square of $1+\hat{u}(\bar{q})$, which can be expressed
as: \begin{eqnarray*}
\hat{J}_{-1/2}(\bar{q}) & \equiv & 1-\frac{1}{2}\hat{u}(\bar{q})+\frac{3}{8}\hat{u}^{2}(\bar{q})\approx(I+\hat{u}(\bar{q}))^{-1/2}\\
 & = & A_{-1/2}(\bar{q})+\bigl[B_{-1/2}(\bar{q})P\bigr]^{(2)}+\bigl[C_{-1/2}(\bar{q})P\bigr]^{(4)}\end{eqnarray*}
 with \begin{eqnarray}
\begin{cases}
A_{-1/2}(\bar{q})=1-\frac{3}{2^{9}}\left(\bar{N}^{(2)''}(\bar{q})\right)^{2}\\
B_{-1/2}(\bar{q})=-\frac{1}{4}\left(1+\frac{1}{16}\bar{N}^{(2)''}(\bar{q})-\frac{3}{8}\frac{\left(\bar{N}^{(2)'}(\bar{q})\right)^{2}}{\bar{N}^{(2)}(\bar{q})}\right)\bar{N}^{(2)}(\bar{q})\\
C_{-1/2}(\bar{q})=\frac{3}{32}\left(\bar{N}^{(2)}(\bar{q})\right)^{2}\end{cases}\end{eqnarray}

\subsection{Validity of the approximation used to calculate the operator $(I+\hat{u}(\bar{q}))^{1/2}$:}

The validity of the truncation made on the expansion of the square
root of the overlap kernel $\hat{J}_{1/2}(\bar{q})$, Eq. (\ref{expJ1/2}),
is checked here, by estimating the extent to which the relation $(I+\hat{u}(\bar{q}))^{1/2}(I+\hat{u}(\bar{q}))^{1/2}=(I+\hat{u}(\bar{q}))$
is satisfied. In other words, using the expression of $(I+\hat{u}(\bar{q}))^{1/2}$
given in Eq. (\ref{expJ1/2}), the residual $R(\bar{q})$ is defined
by the relation \[
(I+\hat{u}(\bar{q}))^{1/2}(I+\hat{u}(\bar{q}))^{1/2}=(I+\hat{u}(\bar{q}))+R(\bar{q})\]
 with \[
R(\bar{q})=-\frac{1}{8}(\hat{u}(\bar{q}))^{3}+\frac{1}{64}(\hat{u}(\bar{q}))^{4}\]
 Consistently with the truncations made in section \ref{SCIM}, the
residual $R(\bar{q})$ is restricted to its lower orders in its development
in symmetric ordered products. Thus $R(\bar{q})$ is \[
R(\bar{q})=A(\bar{q})+\frac{1}{2}\bigl[C(\bar{q})\bar{N}^{(2)}(\bar{q})P\bigr]^{(2)}\]
 Neglecting the derivatives $\left(\bar{N}^{(2)}(\bar{q})\right)^{(p)}$
for $p>2$, after tedious but straightforward calculations using Eqs.
(\ref{SOPO2}) and (\ref{eq:SOPO3}), the coefficients $A(\bar{q})$
and $C(\bar{q})$ are \begin{eqnarray*}
A(\bar{q}) & = & \frac{-3}{2^{11}}\left(\bar{N}^{(2)''}(\bar{q})\right)^{3}+\frac{73}{2^{18}}\left(\bar{N}^{(2)''}(\bar{q})\right)^{4}\\
C(\bar{q}) & = & \frac{-1}{2^{9}}\left(9\left(\bar{N}^{(2)''}(\bar{q})\right)^{2}+32\frac{\bar{N}^{(2)''}(\bar{q})\left(\bar{N}^{(2)'}(\bar{q})\right)^{2}}{\bar{N}^{(2)}(\bar{q})}\right)\\
 & + & \frac{5}{2^{13}}\left(34\frac{\left(\bar{N}^{(2)''}(\bar{q})\right)^{2}\left(\bar{N}^{(2)'}(\bar{q})\right)^{2}}{\bar{N}^{(2)}(\bar{q})}+\left(\bar{N}^{(2)''}(\bar{q})\right)^{3}\right)\end{eqnarray*}

The smaller the residual $R(\bar{q})$, and by extension $A(\bar{q})$
and $C(\bar{q})$, the best the approximation. In particular, we find
$A(\bar{q})=C(\bar{q})=0$ for $\bar{N}^{(2)''}(\bar{q})=0$. Similar
conclusions are found for $\hat{J}_{-1/2}(\bar{q})$.


\begin{thebibliography}{35}
\bibitem{Reinhard87} P.-G.Reinhard, K.Goeke, Rep. Prog. Phys. \textbf{50},
1 (1987).

\bibitem{Sabbey07} B. Sabbey, M. Bender, G.-F. Bertsch and P.-H.
Heenen, Phys. Rev. C \textbf{75}, 044305 (2007).

\bibitem{Bender08} M. Bender and P.-H. Heenen, Phys. Rev. C \textbf{78},
024309 (2008).

\bibitem{Rodriguez07} T.R. Rodriguez and J.L. Egido, Phys. Rev. Lett.
\textbf{99}, 062501 (2007).

\bibitem{Schwerdtfeger09} W. Schwerdtfeger et al. Phys. Rev. Lett.
\textbf{103}, 012501 (2009).

\bibitem{Bertsch07} G.-F. Bertsch et al., Phys. Rev. Lett. \textbf{99},
032502(2007).

\bibitem{Delaroche10} J.P. Delaroche et al., Phys. Rev. C \textbf{81},
014303 (2010).

\bibitem{Berger84} J.-F. Berger, M. girod, and D. Gogny, Nucl. Phys.
A \textbf{428}, 23c (1984).

\bibitem{Goutte05} H. Goutte, J.-F. Berger, P. Casoli, and D. Gogny,
Phys. Rev. C \textbf{71}, 024316 (2005).

\bibitem{Didong73} M. Didong, et al., Phys. Rev. C \textbf{14}, 1189
(1973).

\bibitem{Muther77} H. Muther et al., Phys. Rev. C \textbf{15}, 1467
(1977).

\bibitem{Pomme93} S. Pomme et al., Nucl. Phys. A \textbf{560}, 689
(1993).

\bibitem{Pomme94} S. Pomme et al., Nucl. Phys. A \textbf{572}, 237
(1994).

\bibitem{Vives00} F. Vives, F.-J. Hambsch, H. Bax, and S. Oberstedt,
Nucl. Phys. A \textbf{662}, 63 (2000).

\bibitem{Nix84} J.R. Nix et al., Nucl. Phys. A \textbf{424}, 239
(1984).

\bibitem{Scheuter84} F. Scheuter et al., Phys. Lett. B \textbf{149},
149 (1984).

\bibitem{Kolomietz01} V.M. Kolomietz, S.V. Radionov, and S. Shlomo,
Phys. Rev. C \textbf{64}, 054302 (2001).

\bibitem{Davies77} K.T.R. Davies et al., Phys. Rev. C \textbf{16},
1890 (1977).

\bibitem{Carjan86} N. Carjan, A.J. Sierk, and J.R. Nix, Nucl. Phys.
A \textbf{452}, 381 (1986). 


\bibitem{Nadtochy07} P.N. Nadtochy, A. Keli\'{c}, and K.-H. Schmidt,
Phys. Rev. C \textbf{75}, 064614 (2007).

\bibitem{Borunov08} M.V. Borunov, P.N. Nadtochy, and G.D. Adeev,
Nucl. Phys. A \textbf{799}, 56 (2008).

\bibitem{Holzwarth72} G. Holzwarth, Nucl.Phys. A \textbf{185}, 268
(1972).

\bibitem{RingSchuck80} P. Ring and P. Schuck, The Nuclear Many Body
Problem (Springer-Verlag, New-York, 1980).

\bibitem{Flocard75} H. Flocard and D. Vautherin, Phys. Lett. B \textbf{55},
259 (1975).

\bibitem{Bonche90} P.Bonche, J.Dobaczewski, H.Flocard, P.-H. Heenen,
and J.Meyer, Nucl. Phys. A \textbf{510}, 466 (1990)

\bibitem{Haider89}Q. Haider, D. Gogny, and M. S. Weiss, LLNL Tech.
Rep. UCID-21807 (1989).

\bibitem{Haider92} Q. Haider and D. Gogny, J. Phys. G. Nucl. Part.
Phys. \textbf{18}, 993 (1992).

\bibitem{Anguiano01} M.Anguiano, J.L.Egido, L.M.Robledo, Nucl. Phys.
A \textbf{696}, 467 (2001).

\bibitem{Dubray08} N.Dubray, H.Goutte, J.-P.Delaroche, Phys.Rev.
C \textbf{77}, 014310 (2008).

\bibitem{Younes09} W.Younes, D.Gogny, Phys.Rev. C \textbf{80}, 054313
(2009).

\bibitem{Dubray11} N. Dubray, to be submitted.

\bibitem{Robledo08} L. M. Robledo, Phys. Rev. C \textbf{79},021302
(2009).

\bibitem{Bauhoff80}W. Bauhoff, Ann. Phys. \textbf{130}, 307 (1980).

\bibitem{Dietrich10} K. Dietrich, J.-J. Niez and J.-F. Berger, Nucl.
Phys. A \textbf{832}, 249 (2010).

\bibitem{Moya77} E.Moya De Guerra and F.Villars, Nucl.Phys A \textbf{285},
297 (1977). 
\end{thebibliography}
\end{document}